\documentclass[iop, twocolumn]{aastex631}

\usepackage{graphicx}

\usepackage{amsmath}

\received{XXX}
\revised{XXX}
\accepted{XXX}
\submitjournal{JUSTC}

\shortauthors{Guojun Qiao et al.}

\begin{document}

\title{AXPs/SGRs: strange stars with crusts?} 
\correspondingauthor{Guojun Qiao}
\email{gjn@pku.edu.cn}
\correspondingauthor{Renxin Xu}
\email{r.x.xu@pku.edu.cn}
\correspondingauthor{Kejia Lee}
\email{k.j.lee@pku.edu.cn}
\correspondingauthor{Yongquan Xue}
\email{xuey@ustc.edu.cn}
\nocollaboration{1}

\author{Guojun Qiao}
\affiliation{School of Physics, Peking University, Beijing 100871, China}

\author{Lunhua Shang}
\affiliation{School of physics and electronic science, Guizhou Normal University, Guiyang 550025, China}
\affiliation{Guizhou Provincial Key Laboratory of Radio Astronomy and Data Processing, Guizhou Normal University, Guiyang, 550025, China}

\author{Renxin Xu}
\affiliation{School of Physics, Peking University, Beijing 100871, China}

\author{Kejia Lee}
\affiliation{School of Physics, Peking University, Beijing 100871, China}

\author{Yongquan Xue}
\affiliation{CAS Key Laboratory for Research in Galaxies and Cosmology, Department of Astronomy, University of Science and Technology of China, Hefei 230026, China}
\affiliation{School of Astronomy and Space Sciences, University of Science and Technology of China, Hefei 230026, China}

\author{Qijun Zhi}
\affiliation{School of physics and electronic science, Guizhou Normal University, Guiyang 550025, China}
\affiliation{Guizhou Provincial Key Laboratory of Radio Astronomy and Data Processing, Guizhou Normal University, Guiyang, 550025, China}
\author{Jiguang Lu}
\affiliation{National Astronomical Observatories, Chinese Academy of Sciences, Beijing 100101, China}
\author{Juntao Bai}
\affiliation{Xinjiang Astronomical Observatory, Chinese Academy of Sciences, Urumqi 830011, China}

%
\nocollaboration{12}

\begin{abstract}

The emission of Anomalous X-ray Pulsars (AXPs) and Soft Gamma-Ray Repeaters 
(SGRs) is believed to be powered by the dissipation of their strong magnetic 
fields, which coined the name `magnetar'. By combining timing and energy 
observational results, the magnetar model can be easily appreciated. From a 
timing perspective, the magnetic field strengths of AXPs and SGRs, calculated 
assuming dipole radiation, are extremely strong. From an energy perspective, the 
X-ray/soft gamma-ray luminosities of AXPs and SGRs are larger than their 
rotational energy loss rates (i.e., $ L_{\rm X}>\dot E_{\rm rot}$). It is thus reasonable to assume that the 
high-energy radiation comes from magnetic energy decay, and the magnetar model 
has been extensively discussed (or accepted).
However, we argue that: (1) calculating magnetic fields by assuming that 
rotational energy loss is dominated by dipole radiation (i.e.,  $\dot{E}_{\rm rot}\simeq\dot{E}_{\mu}$)) may be 
controversial, and we suggest that the energies carried by outflowing particles 
should also be considered; and (2) the fact that X-ray luminosity is greater 
than the rotational energy loss rate does not necessarily mean that the emission 
energy comes from the magnetic field decaying, which requires further observational 
testing.
Furthermore, some observational facts conflict with the `magnetar' model, such as observations of anti-magnetars, high magnetic field pulsars, and radio and X-ray observations of AXPs/SGRs. Therefore, we propose a crusted strange star model as an alternative, which can explain many more observational facts of AXPs/SGRs.
\end{abstract}
\keywords{stars: magnetic field --- stars: neutron --- pulsars: individual: PSR J1824-0132}

\section{Introduction}

Neutron stars are typically classified as pulsars and are known for their strong magnetic fields.
To produce the observed pulses, pulsars are generally thought to dissipate their rotational energy through dipole radiation to fuel the emission of radiation pulses. Therefore, the magnetic field of a pulsar can be readily calculated by equating the magnetic dipole radiation energy to the rotational energy. 
For the pulsar magnetosphere, if its inner gap has the maximum potential \citep{Ruderman1975} and the current flow has the Goldreich-Julian charge density \citep{Goldreich1969}, then {the current flow} will carry an energy flow, which equals to the magnetic dipole radiation luminosity.  
The radiation must be powered by the current flow, so the luminosity of radiation observed at all wavelengths must be less than the rotational energy loss rate. This is what we had been observed in radio or most gamma-ray pulsars. These pulsars are named as rotation-powered neutron stars.

On the other hand, the rotational energy of {AXPs and SGRs} are insufficient to supply the radiation, i.e., the high energy luminosity $L_{\rm X}$ is larger than the rotational energy loss rate $\dot E_{\rm rot}$. This suggests their emissions are powered by alternative channels instead of the rotational energy. 
To solve the energy problem, several models are proposed: 1) the magnetar model, where the radiation comes from the decay of super strong magnetic field \citep{Duncan1992,Thompson1995,Kaspi2017,Esposito2021}, 
and 2) the fossil accretion disk model, where the radiation is accretion powered \citep{Chatterjee2000}. 
Beside of these models, some other models include strange star models \citep{Cheng1998a,Alcock1986,Haensel1986,Glendenning1992,Pines1985,Geng2021} and 
wind braking models, where the radiation energy comes from particle emission from the stellar surface \citep{1999ApJ...525L.125H,2000ApJ...535L..51Z,2013ApJ...768..144T}.

Magnetars have been widely studied due to their unique and extreme characteristics in radio, X-ray, and soft gamma-ray bands, which exhibit various phenomena, including X-ray short bursts, large outbursts, giant flares, repeated soft gamma-ray bursts \citep{Kaspi2017,Esposito2021,De2020,Torne2020,Esposito2020}, transient radio emission\citep{2006Natur.442..892C}, fast radio bursts \citep{Margalit2020,Dai2020,2020ApJ...902L...2B,Katz2020,Wang2020,2023SciA....9F6198Z,2024ApJ...965...87I} and interesting timing behaviors \citep{Halpern2010,Halpern2012,Gotthelf2020}.
Here we point out: (a) If an accretion process takes place after a supernova explosion and a crust can be formed around a strange star, one can expect that ion particles in the polar cap regions will be formed. 
This magnetosphere of the strange star would be very different from that of radio pulsar \citep{Cheng1998a,Dai1998,Usov2001,Xu2006}. (b) In the polar cap regions, the gravitational force of the ions is balanced by electric force; and the balance is unstable, so the ions are easy to fall down to the strange star surface along magnetic tube in polar cap regions and the phase transition process will occur, which provides enough energy to push the ions in the crust to flow out freely. (c) Owing to the observed luminosity being higher than the rotational energy loss of the star for AXPs/SGRs, one cannot calculate the magnetic field ($B$; $B_{\rm s}$ for surface magnetic field; {in units of Gauss, G}) with $P$ and $\dot{P}$ using the method of $B_{\rm s} = 3.2\times10^{19}(P \dot{P})^{1/2}$ under the scenario of dipole magnetic radiation at all, where $P$ is the spin period {(in units of second, s)} and $\dot{P}$ is the period derivative {(in units of s~s$^{-1}$)}.

It is evident that any theoretical models must withstand the test of observations. Currently more and more observations become available, especially the radio emission phenomenon. Up to now, more than 30  sources are listed as magnetars \citep{2014ApJS...212..6O} (see the McGill Online Magnetar Catalog\footnote{http://www.physics.mcgill.ca/$\sim$pulsar/magnetar/main.html}),  including 16 SGRs (12 confirmed and 4 candidates), 14 AXPs (12 confirmed and 2 candidates), and 6 with pulsed emission observed in the radio band,  
25 of which have period and period derivative measurements. 
Magnetar-like activities have been recorded from other classes of isolated neutron stars, such as the high-B radio pulsars \citep{2016ApJ...829L..21A,2023ApJ...952..120H} and the source at the center of the supernova remnant RCW 103 \citep{2016ApJ...828L..13R}.

Instead of the magnetar model, we propose a scenario of a strange star with a crust, which is a crusted strange star {with} an accretion disk, to explain the observations. 
In this model, the persistent X-ray emission, burst luminosity, and spectra of AXPs/SGRs can be understood naturally. 
The radio-emitting AXPs, which challenge the magnetar model, can also be explained by this model.
{Besides the evidence for bare strange stars~\citep{1999ApJ...522L.109X,2002ApJ...570L..65X}, crusted strange stars may behave differently during the phase of X-ray bursts, as of millihertz quasi-periodic oscillations~\citep{2023MNRAS.525.2054L}. Nevertheless, an effort to verify the model of strange stars with/without a crust should be encouraged in the future. More evidence for strange stars could be obtained if cold strange matter is in a solid state~\citep{2003ApJ...596L..59X}, e.g., star-quaked induced repeating fast radio bursts~\citep{2018ApJ...852..140W}.}

The paper is organized as follows.
In Section 2, we summarize the observations that in tension with the magnetar model.
In Section 3, we take into account the energy loss processes of particle outflow and recalculate the magnetic field strength of AXPs/SGRs, which indicates that the strong magnetic field was over estimated.
In Section 4, we propose a model of a strange star with crust to understand the X-ray and radio observations of AXPs/SGRs. 
In Section 5, we present the discussions and conclusions.

\section{Observations that are in tension with the magnetar model}
\subsection{Observations of anti-magnetars}

{That of ``$L_{\rm X} > \dot{E}_{\rm rot}$''} does not mean that the super-strong magnetic field really exists in the magnetar.
Anti-magnetar observations in both X-ray and radio bands challenge the super-strong magnetic field assumption.
{Central compact objects (CCOs) are isolated neutron stars in supernova remnants (SNRs), with thermal X-ray emission dominating the emission, which are popularly supposed to be anti-magnetars, since their characteristic ages are much larger than the typical ages of SNRs. An anti-magnetar may apparently have a weak magnetic field but an extremely strong one inside.} 
Halpern and Gotthelf provided the first timing measurements of {CCOs (e.g., PSR J1852+0040) in SNRs} (see Table \ref{tbl_1}) by using the archival data of XMM-Newton and Chandra \citep{Halpern2010}.
Their measurement shows that $P$ and $\dot{P}$ of PSR J1852+0040 are 
$105\,\rm{ms}$ and $8.68\times 10^{-18}\,\rm{s\,s^{-1}}$, respectively.
This leads to a rather low surface magnetic field of $B_{\rm s} = 3.1\times 10^{10}\rm{G}$, which is two orders of magnitude lower than the magnetic field measured for normal pulsar population. Howevere,
$L_{\rm X}$ and $\dot{E}_{\rm rot}$ of this source are $\sim 5.3\times10^{33}$\,\rm{erg/s} and $\sim 3.0\times 10^{32} \,\rm{erg/s}$, respectively, with $L_{\rm X}$ being about 18 times of $\dot{E}_{\rm rot}$.
These measurements show that although $L_{\rm X}$ is larger than $\dot{E}_{\rm rot}$, $B_{\rm s}$ can be still very low. Therefore, the radiation energy should not come from the magnetic field decay.


\subsection{Dipole field assumption is invalid for the magnetar}
Assuming that the dipole field radiation dominates the spin-down process, the characteristic ages $\tau_{\rm c}$ is $P/(2 \dot{P})$. For nearly all CCOs, the characteristic ages are not consistent with the ages of {associated SNRs} \citep{Halpern2010}.
For example, $\tau_{\rm c}$ of J1852+0040 is 192\,\rm{Myr}, but the age of associated SNR Kes 79 is about 7\,\rm{kyr};  
$\tau_{\rm c}$ of RX J0822$-$4300 ($P = 112 \,\rm{ms}$) is $\tau_{\rm c} >220$\,\rm{kyr}, but the age of associated SNR Puppis A is about 3.7\,\rm{kyr} \citep{2009ApJ...695L..35G}; and $\tau_{\rm c}$ of 1E 1207.4$-$5209 ($P=424\,\rm{ms}$) is $\tau_{\rm c} > 27\,\rm{Myr}$ or from 200\,\rm{kyr} to 900\,\rm{kyr} \citep{Shi2003}, but the age of associated SNR PKS 1209.51/52 is 7\,\rm{kyr} \citep{2007ApJ...664L..35G}. We summarize the characteristic and host SNR ages in Tab.~\ref{tbl_1}.
As the dipole field assumption is {questionable}, the magnetic field calculated from $P$ and $\dot{P}$ assuming dipole field may also be incorrect.

\begin{table*}[htb]
\caption{Parameters of three CCOs: $P$, $\dot{P}$, characteristic age $\tau_{\rm c}$, host object age $\tau_{\rm host}$ and the host name\label{tbl_1}}
\begin{center}
\setlength{\tabcolsep}{0.05pt}
{\begin{tabular}{lccccccr}
\hline
Name &  $P$\,(\rm{ms}) & $\dot{P}$\,(\rm{s/s}) &  $\tau_{\rm c}$\,(\rm{yr})  &  $\tau_{\rm host}\,(\rm{yr})$ & Host object &  Reference\\ 
\hline
1E 1207.4$-$5209     &	424	 & 6.6$\times 10^{-17}$	  &  {$>2.7\times 10^7$}  & 7000 &	SNR PKS 1209.51/52 & \citep{2009ApJ...695L..35G} \\	
PSR J1852$+$0040  & 105   & 8.7$\times 10^{-18}$ &	 {1.92$\times 10^8$}	 &	7000  &	Kes 75 & \citep{Halpern2010}\\		
RX J0822$-$4300     & 112	 & $<$8.3$\times 10^{-15}$  &  {$>$2.2$\times 10^5$} & 3700 & SNR Puppis A	 & \citep{Halpern2010}\\	
\hline 
\end{tabular}}
\end{center}
\end{table*} 

\subsection{Observations of high-$B$ pulsars}

It is generally believed that the observational differences between the normal radio pulsars and magnetars are caused by the difference in their magnetic fields.
However, some observations show that several radio pulsars have apparent `stronger' magnetic fields than those of some AXPs, which may challenge the magnetar model.
Table\,\ref{tbl_2} presents a comparison between four high-$B$ radio pulsars, three AXPs, one SGR, and one CCO.
PSRs J1847$-$0130, J1814$-$1733, J1718$-$3718, and J1846$-$0258 are four high-$B$ pulsars.
Among them, PSRs J1847$-$0130, J1814$-$1733 and J1718$-$3718 have radio emission; PSRs J1847$-$0130 and J1814$-$1733 have no X-ray emission; and PSR J1718$-$3718 has low X-ray radiation.
While PSR J1846$-$0258 is located in Kes 75 and surrounded by a pulsar wind nebula and an SNR shell, 
its magnetic field is very high ($4.9\times10^{13}$\,\rm{G}) and its radio emission has not been detected so far; 
but it is observed to have the AXP-like bursts and magnetar-like transition \citep{2000ApJ...542L..37G,2008ApJ...688..550A}. 
Its X-ray luminosity is smaller than its spin-down energy, i.e., $L_{\rm X}/\dot{E_{\rm rot}} = 0.05$.
SGR 0418+5729 is an SGR with the lowest magnetic field that is lower than the critical magnetic field $B_{\rm c} = 4.414\times 10^{13}$\,\rm{G}. 
1E2259+586, XTEJ1810$-$197, and 1E 1547.0$-$5408 are three AXPs.
Among them, 1E2259+586 is a glitching AXP, and XTE J1810$-$197 and 1E 1547.0$-$5408 are two AXPs with transient radio emission.
PSR J1852+0040 is a CCO and anti-magnetar located in the center of the SNR Kes 75, whose radio emission has not been detected yet.

\begin{table*}[tbp!]
\small
\caption{Comparison between radio pulsars, AXPs and an anti-magnetar\label{tbl_2}}
\begin{center}
\setlength{\tabcolsep}{2pt}
\begin{tabular}{lcccccr}
\hline
Name &  $P$\,(\rm{s}) & $\dot{P}$\,(\rm{s/s}) &  $\tau_{\rm c}$\,(\rm{yr})  &  $B_{\rm s}\,(\rm{G})$ & Notes &  References\\ 
\hline
PSR J1847$-$0130 & 6.7  & 1.3$\times 10^{-12}$ &	 8.2$\times 10^{4}$	 &	9.4$\times 10^{13}$  &	high-$B$ PSR & \citep{2003ApJ...591L.135M}\\
&&&&&&\citep{2004MNRAS.353.1311H}\\
\\
PSR J1718$-$3718  &	3.3	 & 1.5$\times 10^{-12}$ &  3.5$\times 10^{4}$    &   7.4$\times 10^{13}$  &	high-$B$ PSR & \citep{2005ApJ...618L..41K} \\

PSR J1814$-$1733  & 4.0	 & 7.4$\times 10^{-13}$  &  8.6$\times 10^{4}$   &    5.5$\times 10^{13}$ &     high-$B$ PSR & \citep{2000ApJ...541..367C} \\
\\
PSR J1846$-$0258  & 0.324 & 7.1$\times 10^{-12}$  &  7.2$\times 10^{2}$   &    4.9$\times 10^{13}$ &   high-$B$ PSR, no radio &\citep{2000ApJ...542L..37G}\\
&&&&&&\citep{2008ApJ...688..550A}\\
\\
SGR 0418+5729 & 9.08  & {4.0$\times 10^{-15}$} & 3.6$\times 10^{7}$	 &	6.1$\times 10^{12}$  &	magnetar, no radio & \citep{2013ApJ...770...65R} \\	
1E 2259+586  & 6.98	 & 4.8$\times 10^{-13}$ &  2.3$\times 10^{5}$    &   5.9$\times 10^{13}$  &	magnetar, no radio & \citep{2014ApJ...784...37D}\\	
XTE J1810$-$197  & 5.54	 & {8.0$\times 10^{-12}$}  &  1.1$\times 10^{4}$   &    2.1$\times 10^{14}$ &     magnetar, radio &  \citep{2007ApJ...663..497C} \\
\\
1E 1547.0$-$5408  & 2.07	 & 4.8$\times 10^{-11}$  &  {7.0$\times 10^{2}$}   &    3.2$\times 10^{14}$ &    magnetar, radio & \citep{2012ApJ...748....3D} \\
&&&&&&\citep{2021MNRAS.503.1214C} \\
&&&&&&\citep{2021ApJ...907....7I} \\
&&&&&&\citep{2023MNRAS.523.4089S} \\
\\
PSR J1852+0040  & 0.105 & 8.7$\times 10^{-18}$  &  1.9$\times 10^{8}$   &    3.1$\times 10^{10}$ &    ${L_{\rm X}}/{\dot{E}_{\rm rot}}=17.7$, no radio &\citep{Halpern2010} \\
\hline 
\end{tabular}
\end{center}
\end{table*}

\subsection{Radio emission observed from some magnetars}

It is generally believed that there is no radio emission coming from the magnetars, where the photon splitting in strong magnetic field suppress the electron-positron pair production, which is essential to the pulsed radio emission \citep{1998ApJ...507L..55B}.
However, more and more radio emission from magnetars has been observed in recent years.
Until now there are five magnetars that have been observed to have radio radiation, e.g., XTE J1810$-$197, which is the first detected radio-emitting magnetar. Its radio pulsations were observed after its 2003 X-ray outburst and then the radio pulsations disappeared in late 2008 \citep{Camilo2016}; 
in December 2018\citep{2019MNRAS.488.5251L}, its radio pulsation reactivated, being detected between 12--18 days after the X-ray emission was enhanced\,\citep{Gotthelf2020}.
Another AXP 1E 1547.0$-$5408 was also observed to have transient pulsed radio emission following X-ray bursts \citep{2007ApJ...659L..37C}. Similar cases also include SGR~1935$+$2145, that its radio pulsar state is recently discovered \citep{2023SciA....9F6198Z} .
The pulsed radio emission of these two objects is similar to the typical emission from normal radio pulsars.
These observations challenge the general understanding of magnetars.

\section{Do super-strong magnetic fields really exist in AXPs/SGRs?}

As discussed above, all characteristic ages of CCOs do not agree with the ages of the associated SNRs. 
As shown in Table\,\ref{tbl_03}, the same thing happens in the case of magnetars.
The characteristic ages of magnetars and their host object ages are very different.
If we believe that the ages of SNRs are correct in most situations, then we should ask why the characteristic ages of CCOs and these ``magnetars'' are significantly different from their associated SNRs. Clearly, the characteristic ages are incorrectly estimated by assuming that the rotational energy loss equals dipole magnetic field radiation.
In the $P-\dot{P}$ diagram the AXPs and SGRs are well located at the strong magnetic field area\,\citep{Halpern2010}.
These derived magnetic fields are based on the assumption that 
the rotational energy loss equals dipole magnetic field radiation. As the dipole radiation assumption is problematic in estimating the age, one would naturally be cautious about the magnetic field estimated with the same assumption.

There is also a logical issue in estimating magnetic field in the magnetar models. In the magnetar model, the X-ray luminosity is assumed to transform from the magnetic energy to explain that the rotational energy loss is smaller than the observed X-ray luminosity in AXPs, i.e., $\dot{E}_{\rm rot}\ll L_{\rm X}$. 
Therefore, it is by-definition no valid to estimate the magnetic field with the assumption that the rotational energy loss equals dipole magnetic field radiation, i.e. $\dot{E}_{\rm rot} = \dot{E}_{\mu}$. Magnetars with a super-strong magnetic field of $>10^{13}$\rm{G} {are} shown in Table\,\ref{tbl_04}. For stars with $\dot{E}_{\rm rot}< L_{\rm X}$, the magnetic field estimation is problematic.

In the later part of the paper, we propose that a super-large particle emission can help us understand the above mentioned observations of AXPs and SGRs. 
Below we first discuss rotational energy loss through particle stream.
\begin{table*}[htb]
\caption{Parameters of some ``magnetars''\label{tbl_03}}
\begin{center}
\setlength{\tabcolsep}{1.pt}
{\begin{tabular}{lccccccr}
\hline
Name &  $P$\,(\rm{ms}) & $\dot{P}$\,(\rm{s/s}) &  $\tau_{\rm c}$\,(\rm{yr})  &  $\tau_{\rm host}\,(\rm{yr})$ & Host object &  References\\ 
\hline	
&&&&&&\citep{1981ApJ...246L.127H}\\
1E 2259+586    & 6.98	 & 4.84$\times 10^{-13}$  &  2.3$\times 10^{5}$ & 1.7$\times 10^{4}$ &  SNR:CTB 109	 & \citep{1992PASJ...44....9I} \\
&&&&&&\citep{2001ApJ...550..397M}\\
&&&&&&\citep{2014ApJ...784...37D}\\
\\
SGR 1806$-$20  & 7.55  & 4.95$\times 10^{-10}$ &	{2.4$\times 10^{2}$}	 &	(3--5)$\times 10^{6}$  &	MSC &  \citep{2007ApJ...654..470W}\\
\\
SGR 1900+14     &	5.20	 & 9.2$\times 10^{-11}$	  &  {9.0$\times 10^{2}$}  & {$10^6$--$10^7$}  & SNR:CTB 109 & \citep{2000ApJ...533L..17V}\\
&&&&&&\citep{2006ApJ...653.1423M}\\
\hline 
\end{tabular}}
\end{center}
\end{table*} 

\begin{table*}[htb!]
\small
\centering
\caption{Parameters of some AXPs (obtained from the McGill Online Magnetar Catalog \citep{2014ApJS...212..6O}; $L_{\rm X}$ is given for the 2--10 keV band)}
\label{tbl_04}
\resizebox{\textwidth}{32mm}{
\setlength{\tabcolsep}{2pt}
\begin{tabular}{lcccccr}
\hline
Source name &  $P$\,(\rm{s}) & $\dot{P}$\,(\rm{s/s}) & $\tau_{\rm c}$\,(yr) & $B_{\rm s}$\,(\rm{G}) & $\dot{E}_{\rm rot}$\,(\rm{erg/s}) & $L_{\rm X}$\,(\rm{erg/s})\\ 
\hline
CXOU J010043.1$-$721134 & 8.02 & 1.88$\times 10^{-11}$ & 6.8$\times 10^{3}$ & 3.90$\times 10^{14}$ & 1.400$\times 10^{33}$ & 6.50$\times 10^{34}$ \\
4U 0142+61 & 8.69 & 2.00$\times 10^{-12}$ &  6.8$\times 10^{4}$ & 1.30$\times 10^{14}$ & 3.369$\times 10^{11}$  & 1.05$\times 10^{35}$ \\
1E 1048.1$-$5937 & 6.64 & 2.25$\times 10^{-11}$ &  4.5$\times 10^{3}$ & 3.90$\times 10^{14}$ &  3.330$\times 10^{33}$ & 4.90$\times 10^{32}$ \\
1E 1547.0$-$5408 & 2.07 & 4.77$\times 10^{-11}$ &  {6.9$\times 10^{2}$} & 3.20$\times 10^{14}$ &  2.100$\times 10^{35}$ & 1.30$\times 10^{33}$ \\
PSR J1622$-$4950 & 4.33 & 1.70$\times 10^{-11}$ & 4.0$\times 10^{3}$  & 2.70$\times 10^{14}$ & 8.300$\times 10^{33}$ & 4.40$\times 10^{32}$ \\
CXOU J164710.2$-$455216 & 10.61 & 4.00$\times 10^{-13}$ &  4.2$\times 10^{3}$ & 6.60$\times 10^{13}$ & 1.300$\times 10^{31}$ & 4.50$\times 10^{32}$ \\
1RXS J170849.0$-$400910 & 11.01 & 1.95$\times 10^{-11}$ &  9.0$\times 10^{3}$ & 4.70$\times 10^{14}$ & 5.800$\times 10^{32}$ & 4.20$\times 10^{34}$ \\
XTE J1810$-$197 & 5.54 & 7.80$\times 10^{-12}$ &  {1.1$\times 10^{4}$} & 2.10$\times 10^{14}$ & 1.800$\times 10^{33}$ & 2.50$\times 10^{34}$ \\
1E 1841$-$045 & 11.79 & 4.09$\times 10^{-11}$ &  4.6$\times 10^{3}$ & 7.00$\times 10^{14}$ & 9.900$\times 10^{32}$ & 1.84$\times 10^{34}$ \\
1E 2259+586 & 6.98 & 4.80$\times 10^{-12}$ &  2.3$\times 10^{5}$ & 5.90$\times 10^{13}$ & 5.600$\times 10^{31}$ & 1.70$\times 10^{34}$ \\
\hline 	
\end{tabular}}
\end{table*} 

\subsection{Rotational energy loss of neutron stars}
The rotational energy loss of pulsar can be written as:
\begin{equation} 
   \dot{E}_{\rm rot} =
   I \Omega \dot{\Omega},
\end{equation}
where $I = \frac{2}{5}M R^{2} \approx 10^{45}\,\rm{g\,cm^{2}}$ is the moment of inertia, $M$ (in units of solar mass $M_{\odot}$) and $R\approx 10^6\,\rm{cm}$ are the mass and radius of the pulsar, respectively; 
$\Omega$ and $\dot{\Omega}$ are the angular velocity and its derivative of the pulsar, respectively. 


If we still assume that the rotational energy loss is equal to {the sum of the particle emission, $\dot{E}_{p,r}$, and the energy loss of dipole magnetic radiation, $\dot{E}_{\mu}$, one then has}

\begin{equation}\label{eq2}
 I\Omega|\dot{\Omega}| =  \dot{E}_{p,r} + \dot{E}_{\mu}.
\end{equation}
In the RS model \citep{Ruderman1975}, the maximum rotational energy loss through particle emission can be written as
\begin{equation}
\dot{E}_{p,r} \approx \frac{B^{2}_{s}\Omega^{4}R^{6}}{c^{3}},
\end{equation}
and the dipole magnetic energy loss can be written as
\begin{equation}
\dot{E}_{\mu} \approx \frac{B^{2}_{s} \Omega^{4} R^{6}}{c^{3}}.
\end{equation}
As shown above, the rotational energy loss is limited by the magnetic field. 
However, for AXPs, the observed fact is
\begin{equation}
\dot{E}_{\rm rot} = 4\pi^{2} I \dot{P} P^{-3} \ll L_{\rm X},
\end{equation}
where $L_{\rm X}$ is the observed X-ray luminosity.
To resolve this problem, one way for the magnetar model is to assume that the pulsar has a strong dipole magnetic field and the energy loss of the dipole magnetic field is larger than the rotational energy loss. 
From the discussion above, if the charge density $n$ inside the magnetosphere is the GJ charge density $n_{gj}$ \citep{Goldreich1969} and the inner gap potential $\Delta{V}=\Delta{V_{\rm m}}$ \citep{Ruderman1975}, the rotational energy loss $E_p$ through particle emission and energy carried out by particles also increase, that is $E_p \approx E_{p,r} \approx E_{\mu}$, 
where $E_p$ is the energy carried out by particles (see next section).
May the particle energy loss rate be larger than the rotational energy loss rate? If so, what is this physical situation?
Let us deal with this from the RS model first.

\subsection{Energy carried out by particle emission}

The energy carried out by particle emission can be estimated with the RS model as below. 
The maximum charge-particle flux from one polar cap can be represented as 
\begin{equation}
\dot{N}_{\rm max} = \pi r^{2}_{p} c n_{gj} = \pi r^{2}_{p} c \frac{\Omega\cdot B_{s}}{2 \pi e c}.
\end{equation}
The energy loss by the particles from the two polar caps can be written as 
\begin{equation}
\dot{E}_{p} = 2 \dot{N}_{\rm max} \gamma m_e c^{2},
\end{equation}
where $\gamma = \frac{e \Delta{V_m}}{2 c^{2}}$ is the Lorentz factor.
The maximum potential drop in the inner gap is $\Delta{V_m} = \frac{\Omega B}{2 c}r^{2}_p$ and $r_p = R(R\Omega/c)^{1/2}$ is the polar cap radius of pulsar.
Then, the energy loss rate through particles can be obtained as 
\begin{equation}\label{e8}
\dot{E}_p = \frac{B^{2}_s \Omega^{4} R^6}{2 c^3}.
\end{equation}
This means that under the RS model, the maximum energy loss through particles is equal to the dipole magnetic energy loss rate approximately. In other words, the higher magnetic field will result in greater particle emission.
This is why the magnetars can produce higher emission despite having a much lower rotational energy loss rate $\dot{E}_{\rm rot}$ compared to their X-ray luminosity  $L_{\rm X}$, i.e., they have a super-strong magnetic field.

From the discussion above, to resolve the problem of $\dot{E}_{\rm rot} \ll L_{\rm X}$, increasing the magnetic field is a possible way. 
However, there is another possibility to resolve this problem that the particle flow density is high ($n \gg n_{gj}$) although the magnetic field is normal.

\subsubsection{The normal wind situation}
Let us look at the rotational energy loss through the particle emission. In the magnetosphere of the pulsar, the particles co-rotate with the star due to the existence of a strong magnetic field. 
The premise of the particles co-rotating with the star is that the magnetic energy density is larger than the particle energy density.
If assuming $r_c$ is the co-rotating radius and at which particles move out with an area of $\pi r^{2}_c$, then the particle energy loss rate is, 
\begin{equation}
\dot{E}_p = 2 \pi r^{2}_c c \varepsilon_p,
\end{equation}
where $\varepsilon_p$ is the  particle energy density,
\begin{equation}
\varepsilon_p = \frac{\dot{E}_p}{2 \pi r^{2}_c c}.
\end{equation}
At  the co-rotating radius $r_c$, the magnetic energy density is
 \begin{equation}
 \varepsilon_b = \frac{B^{2}_s R^6}{8 \pi r^{6}_c}.
\end{equation}
Let the particle density equal the magnetic energy density, $\varepsilon_p = \varepsilon_b$, then we have
\begin{equation}\label{e12}
r_c = \left(\frac{c B^{2}_s R^6}{2 \dot{E}_p}\right)^{\frac{1}{4}}.
\end{equation}
Substituting $\dot{E}_p$ in equation\,(\ref{e8}) into the equations above, we obtain $r_c = c/\Omega = R_c$, where $R_c$ is the radius of the light cylinder of the pulsar.
This agrees with the normal estimate, i.e., the particles co-rotating with the pulsar within the light cylinder radius in the magnetosphere.

The rotational energy loss carried out by the particles from one polar cap of the pulsar is
\begin{equation}
\dot{E}_{p,r} = \dot{N}_s \gamma_s m_e \nu^{2}_c,
\end{equation}
where $\dot{N}_s$ is the total particles flowing out from one pole, and
$\nu_c$ is the co-rotating velocity. Near the light cylinder, $\nu_c = c$.
If there is no acceleration outside of the gap, from the conversion of energy we can obtain $\dot{N}_s \gamma_s = \dot{N}_p \gamma_p$.
Then, the rotational energy loss carried out by the
particles can be written as 
\begin{equation}
\dot{E}_{p,r} \approx \frac{B^{2}_s \Omega^4 R^6}{c^3},
\end{equation} 
which shows that, in the RS model, the rotational energy loss induced by the particle emission is the same as that by the dipole magnetic radiation. 
Increasing the particle energy loss rate is the same as increasing the magnetic field.

\subsubsection{The strong wind situation}
When the magnetic field is not strong enough and the energy carried out by particles is large enough, i.e., $\dot{E}_p \gg \dot{E}_{\mu}$, then the magnetic fields and ages of AXPs will be calculated differently from that calculated  simply assuming $\dot{E}_{\rm rot} = \dot{E}_{\mu}$.
If the energy loss rate carried out by particles is equal to or larger than the observed X-ray luminosity, we can write
\begin{equation}
\dot{E}_p = 2 \dot{N_p} \gamma_p m_e c^{2} = \xi L_{\rm X},
\end{equation} 
where $\xi = \dot{E}_p/L_x$ is a coefficient, and $\dot{N_p}$ is the number of particles flowing out of one pole per unit time.  
The rotational energy loss carried out by particles from two poles is 
\begin{equation}
\dot{E}_{p,r} = \dot{N_p} \gamma_p m_e \nu^{2}_c,
\end{equation}
where $\nu_c$ is the co-rotating velocity
\begin{equation}
\nu_c \approx \Omega r_c = \Omega (\frac{c B^{2}_s R^{6}}{2  \xi L_x})^{\frac{1}{4}}.
\end{equation}
If we assume 
\begin{equation}
\dot{E}_{p,r} = \dot{E}_{rot} = 4 \pi^2 I \dot{P} P^{-3},
\end{equation}
we obtain the magnetic field from the equations above,  
\begin{equation}
B_s = \frac{8\sqrt{2} \pi^2 I \dot{P} c^{\frac{3}{2}}} {( \xi L_x)^{\frac{1}{2}} \Omega^2 P^3 R^3},
\end{equation}
or 
\begin{equation}\label{eq20}
B_s =  (1.5\times 10^{25}\,\rm{Gauss}) I_{45} \dot{P} P^{-1} \xi^{-\frac{1}{2}} L^{-\frac{1}{2}}_{x,35}.
\end{equation}
From the equations above, we can obtain

\begin{align}
\begin{split}
\frac{n}{n_{gj}} &= \frac{\xi L_x}{2 \pi r^{2}_p c^3 \gamma m_e n_{gj}} \\
                 &= \frac{4.3 \times 10^{23} \xi L_{\rm X,35} P^{2}}{\pi^2 B_s \gamma}.
\end{split}
\end{align}
Jessner et al. (2001) showed that for neutron stars with lower surface temperature from $T_{\rm surf} \approx 5\times 10^{5}$\,\rm{K} to $7.5 \times 10^{5}$\,\rm{K} and the work function for standard pulsars, they obtained $n/n_{gj}$ of $10^{6}$ to $10^8$ or more \citep{2001ApJ...547..959J}.
For $\gamma = 10^{8}$, $B_s = 10^{12}$ and $\xi = 100$, we obtain $n/n_{gj} \approx 4.4\times 10^{6} L_{\rm X,35}$.
Therefore, for $\xi = 100$, $P = 1$\,\rm{s} and $\dot{P} = 10^{-12}$, one can obtain $B_s = 1.5\times10^{12}$\,\rm{G} from Equation (20).

\begin{table*}[htb!]
\centering
\caption{Comparison between $B_{s}$ derived from $B_s = 3.2\times 10^{19}(P/\dot{P})^{1/2}$ and $B^a_s$ estimated with Equation\,(\ref{eq20}) ($\xi = 100$ adopted and $\xi_b = B^a_s/B_s$)}
\label{tbl_05}.
\setlength{\tabcolsep}{15pt}
\begin{tabular}{lccr}
\hline
Source name &  $B_s$\,(\rm{G}) & $B^a_s$\,(\rm{G}) & $\xi_b$\\ 
\hline
CXOU J010043.1$-$721134  & 3.900$\times 10^{14}$ & 4.361$\times 10^{12}$   &   0.011\\
4U 0142+61   & 1.300$\times 10^{14}$ & 3.369$\times 10^{11}$   &   0.003\\
CXOU J164710.2$-$455216  & 6.600$\times 10^{13}$ & 8.430$\times 10^{11}$   &   0.013\\
1RXS J170849.0$-$400910 & 4.700$\times 10^{14}$ & 4.099$\times 10^{12}$   &   0.009\\
XTE J1810$-$197  & 2.100$\times 10^{14}$ & 4.224$\times 10^{12}$   &   0.020\\
1E 2259+586  & 5.900$\times 10^{13}$ & 2.502$\times 10^{12}$  &   0.042\\
\hline 	
\end{tabular}
\end{table*} 

It is found that if a strong wind situation is taken into account, the estimated magnetic fields of some AXPs with $L_{\rm X}\gg\dot{E}_{\rm rot}$ are smaller than that derived from $B_s = 3.2\times 10^{19}(P \dot{P})^{1/2}$.
In Table\,\ref{tbl_05},  a comparison between $B_s = 3.2\times 10^{19}(P/\dot{P})^{1/2}$ and the magnetic fields estimated with Equation\,(\ref{eq20}) is presented. The estimated magnetic fields with Equation\,(\ref{eq20}) for CXOU J010043.1$-$721134, 4U 0142+61, and 1E 1841$-$045 are $4.36\times 10^{12}$\,\rm{G}, $3.37\times10^{11}$\,\rm{G}, and $1.83\times10^{12}$\,\rm{G}, respectively, all being smaller than that estimated by $B_s = 3.2\times 10^{19}(P \dot{P})^{1/2}$. But where does the energy of the particles come from? This is what we will discuss in the next section.

\section{A crusted strange star model}

Instead of the magnetar model, we propose a strange star model, which is a crusted strange star {with} an accretion disk. 
In this model, the persistent X-ray emission burst luminosity and spectra of AXPs and SGRs can be understood naturally, and  
the radio emission of AXPs/SGRs that challenges the magnetar model can also be understood.

\subsection{Strange star with a crust: X-ray and radio emission}

To find the energy source of particle emission, a strange star with a crust has been investigated. 
It is assumed that the crust of a strange star would be formed after a supernova explosion
\citep{Alcock1986}. For a typical strange star with a radius of {$1\times 10^{6}$ cm} and a mass of $1.4\ M_{\odot}$, the mass of the crust is usually in the range of $10\times 10^{-7} - 10\times 10^{-5} M_{\odot}$, and the thickness of the  crust is $\sim10^{4}$\,\rm{cm} \citep{Huang1997}.
The distance from the crust bottom to the strange star surface is $\sim200$\,\rm{fm}.
This type of strange star cannot be observed as radio pulsars due to the existence of crust, but bare strange stars can be observed as radio pulsars \citep{Xu2001}.  
Therefore, we suggest that a strange star with a crust could be used to explain the observed X-ray and radio emission of AXPs/SGRs.

\begin{figure*}[htb]  
    \includegraphics[width=\textwidth]{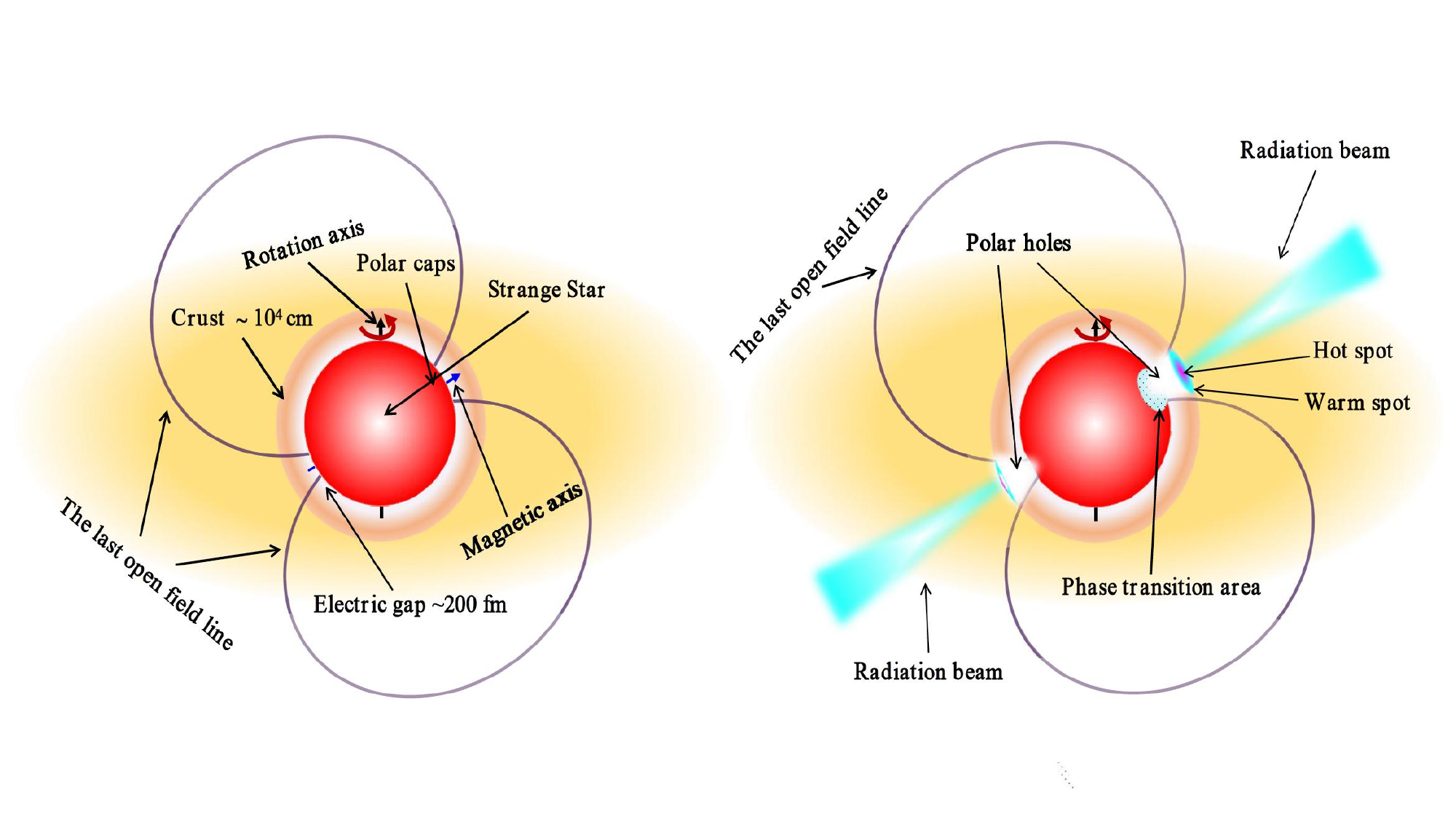} 
   \caption{Schematic illustration of {a crusted strange star with} an accretion disk. 
   Around the strange star there is a crust: in the normal case, we can observe radiation in X-ray bands; and in the super flare, high energy particles flow out from the pole cap regions, so there are two polar holes formed in the crust and we can observe radio emission from polar cap regions.}
   \label{fig1}
\end{figure*}

In the persistent X-ray emission state, the strange star accretes matter from its environment.
The crust on the polar cap surface will become heavier and heavier with a large number of accreted matter accumulated on the surface of the polar cap.  
The crust may finally break to lead to a super flare due to the release of a large number of magnetic energy and plenty of electron/positron pairs on a very short timescale.
In this case, two holes will form in the crust, the polar cap of the strange star will become bare, and the electron/positron pairs in the polar cap surface of the strange star will be accelerated and outflow along the magnetic field lines to generate the radio emission from the bared regions.
With the strange star continuously accreting matter from its environment, the polar holes will be gradually filled by the accreted matter and then the strange star will become radio quiet \citep{2010arXiv1005.3911Q}. A schematic illustration of a strange star with a crust is shown in Figure\,\ref{fig1}.

\subsection{Energy source of emitting particles}

To form a crust for the strange star, an accretion process must take place. 
For a magnetized star, the accreted matter falls onto the surface through the magnetic tube along the open field lines.
The magnetosphere of this kind of star may be different from that of radio pulsars, i.e., ion particles in the polar cap region would be formed and are supported by electric force.
Usually, in the polar cap, compared to the electromagnetic force $F_e$, the gravitational force $F_g$ can be neglected due to
\begin{equation}
\frac{F_g}{F_e} = \frac{G M A m_p c}{\Omega B R^{3} Z e},
\end{equation}
where $A$ and $Ze$ are the mass number and charge number of the ions, respectively; 
$m_p$ is the mass of a proton, and $G$ is the gravitational constant. 
For a dipole magnetic field, $B = B_0 (R/r)^3$ is used, {and $\Omega = 2 \pi/P$}.
If adopting $M = 1 M_{\odot}$, $B_0 = 10^{12}$\,\rm{G}, $P = 1$\,\rm{s} and $R = 10^{6}$\,\rm{cm}, $F_g/F_e $ is $2 \times 10^{-9}(A/Ze)$, 
which means that the electromagnetic force can support any ions to form a magnetosphere.

For a rotating magnetized star, the balance of the magnetosphere depends on the charge density. 
If there is any disturbance, the balance will break down easily, especially for ions. 
If the ions are ionized above the polar cap regions, the charge density would be changed, which means $\rho -\rho_{gj}$ being nonzero.
In this case, the ions will move along the magnetic field lines, and the force on them can be written as \citep{1975ApJ...198..683M,1977ApJ...216..865C}:
\begin{equation}
\frac{dE_{\mu}}{dZ} = 4 \pi(\rho-\rho_{gj}),
\end{equation}
where $\rho_{gj} = \frac{\Omega B}{4 \pi c}$ is the charge density in the GJ magnetosphere, and
$E_{\mu}$ is the electric field component along $B$ which also defines the $Z$ direction. 
This means that when the balance is broken, ions in the magnetosphere will fall onto the crust. 
When the accreted matter falls onto the surface of the crust above the polar cap and reaches a critical mass $\Delta{m}$, the crust above the polar cap will be broken to make the polar cap of the star bare. 
The baring process of the crust is presented as follows.

As discussed above, a crusted strange star cannot be observed as radio pulsars, but bare strange stars can be observed as radio pulsars.
We assume that the crust above the polar cap is broken to bare due to falling matter continuously accumulated on the crust surface, and the bare area is $A_{b} = \eta A_{p}$, where $\eta <=1$ is a coefficient, and $A_ {p}$ is the area of the polar cap, which can be written as 
\begin{equation}
A_{p} = \pi r_{p}^2 = \pi  (R (\frac{2\pi R}{P c})^{\frac{1}{2}})^{2} =\frac{2\pi^2R^3}{Pc},
\end{equation}
where $r_{p}$ is the radius of {the polar cap.}
The energy density gravitationally exerted by the falling matter $\Delta{m} $ on the crust is
\begin{equation}
E_{exert} \simeq \frac{\Delta{m} g}{A_{b}},
\end{equation}
where $g = \frac{GM}{R^2}$, and $M=1.4M_\odot$ and $R=10^6\rm{cm}$ are the mass and radius, respectively.

\begin{figure*}[htb] 
   \centering
\includegraphics[width=0.46\textwidth]{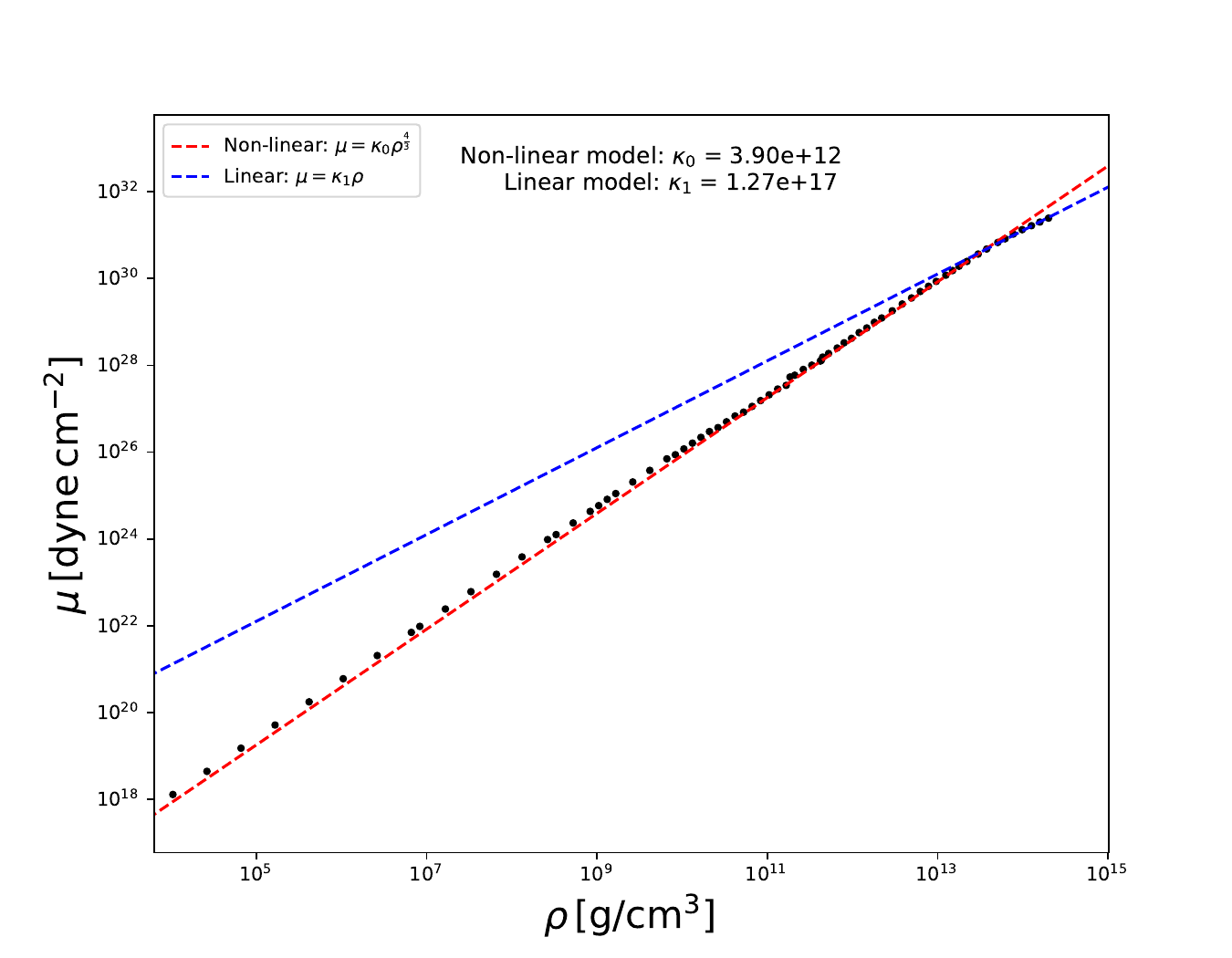} 
\includegraphics[width=0.46\textwidth]{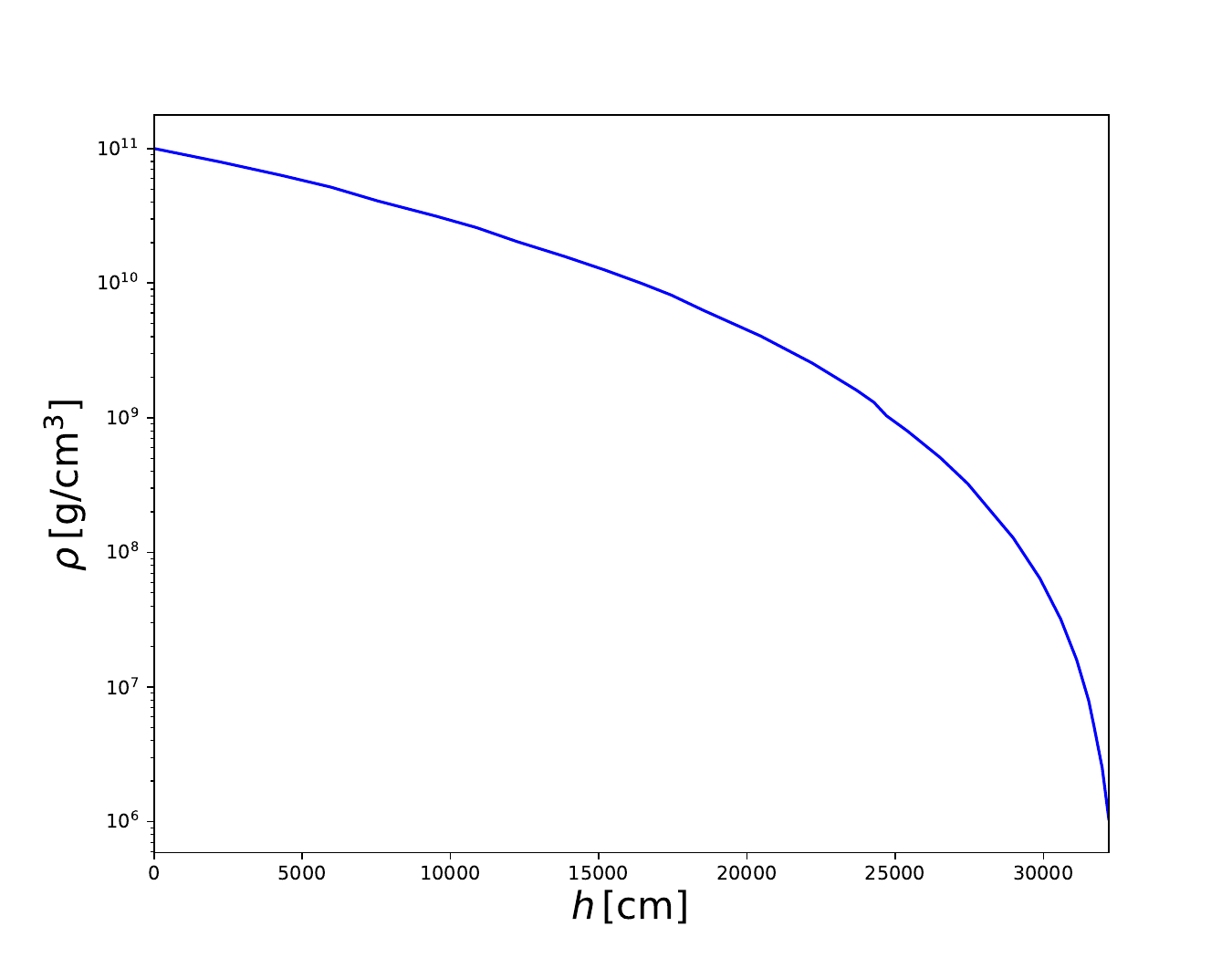} 
   \caption{(Left) Change of $\mu$ with $\rho$. The blue line is a linear model of $\mu \simeq 1.27\times10^{17} \rho$, and the red line is non-linear model of $\mu \simeq 3.9\times 10^{12} \rho^{\frac{4}{3}}$. (Right) Change of crust  density $\rho$ with the height from the surface of star. $h=0$ represents the base of the crust. $\rho$ was obtained from the numerical solution of \citep{Baym1971}. }
   \label{mu}
\end{figure*}

We assume that the density at the base of the crust is $\rho_b = 4\times 10^{11} g/cm^3$, then the maximum elastic energy density of the crust is
\begin{equation}
E_{elast} = \int \theta d\mu \simeq 8.63\times 10^{27} \,\rm{dyne\,cm^2},
\end{equation}
where $\mu$ is the shear-modulus profile in the crust \citep{Haskell2006}, and $\theta \sim10^{-3}$ is the shear angle of the crust \citep{Cheng1998b}.
As shown in the left panel of Figure\,\ref{mu}, we fit the change of $\mu$ with $\rho$, and find that $\rho>10^{13}\,\rm{g\,cm^{-3}}$, $\mu \simeq 1.27\times10^{17} \rho$, while $\mu \simeq 3.9\times 10^{12} \rho^{\frac{4}{3}}$ for $\rho<10^{13}\,\rm{g\,cm^{-3}}$.
Here, due to the crust density $\rho <=10^{11}\,\rm{g\,cm^{3}}$, $\mu \simeq 3.9\times 10^{12} \rho^{\frac{4}{3}}$ is used in our estimation.

\begin{figure*}[htb] 
   \centering
    \includegraphics[width=0.46\textwidth]{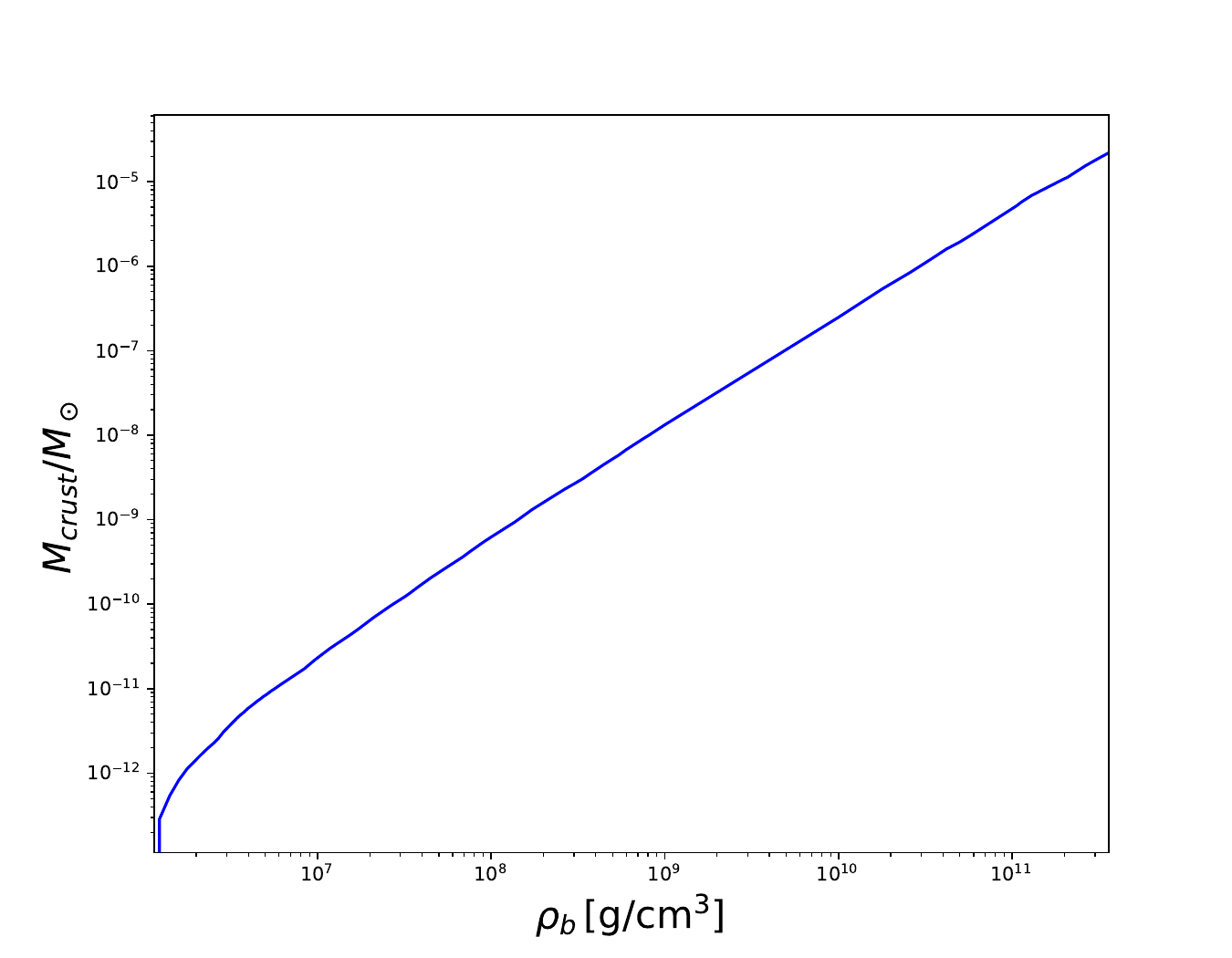} 
    \includegraphics[width=0.46\textwidth]{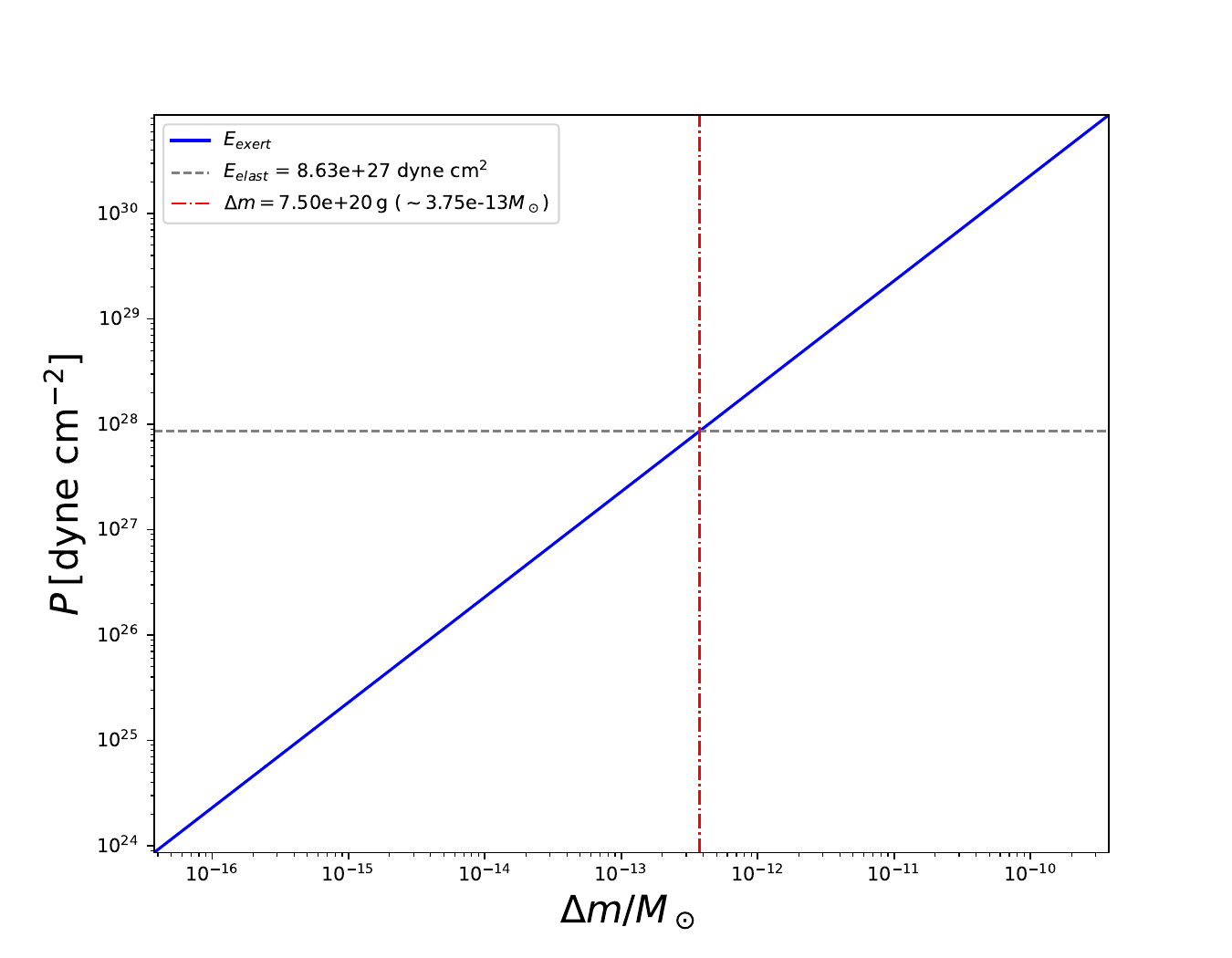} 
\caption{(Left) Change of crust mass with the density $\rho_b$ at the base of the crust, with the surface density of crust being $\rho_t\simeq 10^6 \rm{g\,cm^{-3}}$. (Right) The blue line represents the energy density  gravitationally exerted by the accreted matter $\Delta{m}$ on the crust.
The gray dashed line is the maximum elastic energy limit of the crust of XTE J1810–197 for $\rho_b =4\times 10^{11}\,\rm{g/cm^3}$ and $\rho_t = 10^{6}\,\rm{g/cm^3}$. The red dashed-doted line is $\Delta{m}\simeq 3.7\times 10^{-13}\,M_\odot$.}\label{M_rho_b}
\end{figure*}

When the pressure $E_{exert}$ exceeds the critical stress $E_{elast}$ of the crust, the crust will break, i.e., $E_{exert} > E_{elast}$.
On one polar, the mass that causes the crust to break is
\begin{equation}
M_{c} = \frac{A_b}{A_s} M_{crust}, 
\end{equation}
where $A_s = 4\pi R^2$ is the surface area of the star, and $M_{crust}$ is the mass of crust. 
The crust density $\rho$ changes with the height $h$ from the base of the crust to the surface of the star, shown in the right panel of Figure\,\ref{mu}. 
The density at the base is $\rho_b = 4\times 10^{11} \rm{g\,cm^{-3}}$. 
Then the mass of the crust can be estimated, as shown in the left panel of Figure\,\ref{M_rho_b}. For $\rho_b = 4\times 10^{11} \rm{g\,cm^{-3}}$, $M_{crust} \simeq 2.5\times 10^{-5}\,M_\odot$. 

We assume that the critical mass of accreted matter for crust breaking is $\Delta{m}$,
\begin{equation}
\Delta{m} =\frac{A_b E_{elast}}{g}.
\end{equation}
The interval $t$ between crust-breaking events can be written as
\begin{equation}
t = \frac{\Delta{m}+M_{c}}{\dot{M}},
\end{equation}
where $\dot{M}$ is the accretion rate,
\begin{equation}
\dot{M} = \frac{L_x R}{GM} = \frac{L_x}{gR}.
\end{equation}
Then, the interval $t$ can be estimated with 
\begin{equation}
t = \frac{gR(\Delta{m}+M_{c})}{L_x} \simeq \eta 1.93\times10^{10} (P_1 L_{x,35})^{-1}\,\rm{s},
\end{equation}
where $P_1 = P/1\,\rm{s}$, $L_{\rm X,35} = L_{\rm X}/ 10^{35} \rm{erg\,s^{-1}}$, and $L_{\rm X}$ is the observed X-ray luminosity in quiescent state.
When the crust breaks, the normal matter will have a phase transition to strange matter.
The phase transition will release huge energy $E_{pt}$, then the pulsar enters into an outburst state.
The total outburst energy will reach $E_{pt}$ \citep{Cheng1998b},
\begin{equation}
E_{pt} = (\Delta{m}+M_c) c^2 \frac{30}{931}.
\end{equation}

For XTE J1810–197, as shown in the left panel of Figure\,\ref{M_rho_b}, assuming the base density of crust $\rho_b = 4\times 10^{11} g/cm^3$, the estimated mass of crust $M_{crust} \simeq 10^{-5}\,M_\odot$.
When the accreted matter $\Delta{m}\simeq3.7\times10^{-13}\,M_\odot$,  $E_{exert} =E_{elast}$ (as shown in right panel of Figure\,\ref{M_rho_b}), the crust breaks.
The corresponding mass that induces the crust to break is $M_c\simeq10^{-11}\,M_\odot$. 
In addition, the interval $t$ between crust-breaking events, and the total outburst energy for XTE J1810$-$197 are estimated, $t\sim 15$ years, and the total outburst energy is $\sim 10^{43}\,\rm{erg}$.
These are consistent with the observational interval \citep{2019ApJ...874L..25G,Pintore2019} and total outburst energy of AXPs/SGRs given by \citep{CotiZelati2018}.

\section{Discussions and Conclusions}

We present the observational facts that may be inconsistent with the magnetar model, and propose a new method of estimating the magnetic fields of AXPs and SGRs.
By comparing the characteristic ages $\tau_{\rm c}$ of CCOs with the ages of their associated SNRs, we find that $\tau_{\rm c}$ derived from $P$ and $\dot{P}$ are very different from their ages of associated SNRs.
This implies that the general method of calculating the ages of CCOs with $P$ and $\dot{P}$ is problematic.
In this way, it is unconvincing to calculate $B_s$ of AXPs/SGRs based on the assumption that the rotational energy is dissipated dominantly by the dipole magnetic radiation. As is well known, besides the dipole magnetic radiation, there are particle acceleration and radiation, so the rotational energy loss should equal dipole magnetic radiation plus particle acceleration, as given in Equation \ref{eq2}. If taking this into account, ultra-strong magnetic field is not expected for AXPs/SGRs.

For AXPs and SGRs, there is no link between rotational energy and the dipole magnetic radiation. 
Thus, it may be incorrect to assume that AXPs/SGRs have super-strong dipole magnetic fields.
Here we emphasize that for AXPs/SGRs, {it is} logically inconsistent to estimate the magnetic field based on the assumption that the rotational energy loss equals the dipole magnetic radiation.
We argue that the larger $L_{\rm X}$ may be powered by the strong particle flow, and {present the energy loss processes in the situations of normal and strong winds}, respectively.
It is found that the rotational energy loss through particle emission is the same as that carried out by the dipole magnetic radiation.
Thus, increasing particle energy loss can result in $L_{\rm X}>\dot{E}_{\rm rot}$, which has the same effect as increasing the magnetic field.
Under the strong wind situation, the magnetic fields of several AXPs are recalculated, which are {similar to} normal pulsars.
This means that in the strong wind situation, it is not necessary to employ the super-strong magnetic field to explain $L_{\rm X}>\dot{E}_{\rm rot}$ of AXPs/SGRs.    

Observations are important for testing theories.
Some important observations of AXPs can be well understood in the crusted strange star model, such as the observations of XTE J1810$-$197 \citep{2021PASJ...73.1563E,2022MNRAS.510.1996C,2023ApJ...956...93H}.
After the 2003 outburst of XTE J1810$-$197, it returned to the quiescent state; \citep{2011MNRAS.418..638B} studied the surface emission properties of this source and found that the temperature distribution on the star surface is consistent with the expectations of a dipole magnetic field configuration, which suggests the presence of a magnetized atmosphere.
The X-ray observations and spectrual analysis of this source show that the quiescent emission is composed of two thermal components, which implies that there are two concentric thermal hot spots in the polar cap that may be responsible for the emission of XTE J1810$-$197 \citep{Perna2008,Beloborodov2009,Beloborodov2013,Albano2010,2011MNRAS.418..638B,2019ApJ...874L..25G,Borghese2021}.
The hot footpoint is surrounded by a warm X-ray emitting area. 
As the emitting area of the hot spot is much smaller than that of the warm ring component, it is more likely to find a configuration with parallel field lines in the hot region (and most likely perpendicular to the surface) than in the warm region. 
During the X-ray decay of XTE J1810$-$197, the source became radio-bright\,\citep{2005ApJ...632L..29H} and switched on as a powerful radio pulsar, and the radio pulse appeared to change with time\,\citep{2007ApJ...659L..37C,2024MNRAS.528.3833F}.
The observed radio pulse width is $\sim0.15 P$, and the pulse is almost aligned with the X-ray light curve.
These imply that the radiation location of radio emission in the magnetosphere could be higher than that of X-ray emission\,\citep{Perna2008}. 
These characteristics of the X-ray and radio emission from XTE J1810$-$197 could be well understood with the model of a crusted strange star.

The crust-breaking process and the interval between crust-breaking events are presented.
For XTE J1810--197 with a period of 5.54\,\rm{s}, after the outburst in 2003 \citep{2004ApJ...609L..21I}, its X-ray flux density decreased until it returned to the pre-outburst state in early 2007 \citep{Pintore2019}. On December 8, 2018, the burst event of this source was observed again and re-activated as a radio pulsar \citep{2019ApJ...874L..25G}.
The interval between the burst events of 2003 and 2018 is about 15 years. 
The X-ray luminosity in quiescent state is $L_x\simeq 2.5\times 10^{34}\,\rm{erg\,s^{-1}}$.
With these observations, we estimate the interval between crust-breaking events of XTE J1810--197 is about 15 years by considering the bared area being $\sim 3.4\%$ times the polar cap area.
For the outburst, Francesco et al.(2018) presented the results of the systematic study of all magnetar outbursts through a reanalysis of data acquired in about 1100 X-ray observations \citep{CotiZelati2018}.
They presented the total outburst energy of many AXPs/SGRs from $10^{40}\,\rm{erg}$ to $10^{43}\,\rm{erg}$.
The estimated total outburst energy of XTE J1810--197 in this paper is about $10^{43}\,\rm{erg}$, which is consistent with the results of \citep{CotiZelati2018}.
As discussed above, the observations of AXPs/SGRs can be understood in our model, and the further model tests will be carried out through future observations.

It is surely meaningful if AXPs/SGRs are really strange stars with crusts.
In this era of multi-messenger astronomy that encompasses full electromagnetic waves, neutrinos, cosmic rays, and gravitational waves, it is increasingly important to explore the physics of supra-nuclear matter especially in strong gravity.
Besides pulsar-like compact stars, gamma-ray bursts and fast radio bursts are also relevant to the nature of pulsars, and we expect to solve eventually the equation of the state of extremely dense matter in the coming years.

\begin{acknowledgments}
This work is supported by the National Natural Science Foundation of China (Nos. 12273008, 12025303), the National SKA Program of China (No.2022SKA0130104), the Natural Science and Technology Foundation of Guizhou Province (Nos. [2023]024, ZK[2022]304), the Foundation of Guizhou Provincial Education Department (Nos. KY (2020) 003, KY[2022]137, KY[2022]132), the Academic New Seedling Fund Project of Guizhou Normal University (No. [2022]B18),  the Major Science and Technology Program of Xinjiang Uygur Autonomous Region (No.2022A03013-4). This work uses the data from the FAST. FAST is a Chinese national mega-science facility, built and operated by the National Astronomical Observatories, Chinese Academy of Sciences.

\end{acknowledgments}

\bibliography{sample}{}

\begin{thebibliography}{}
\expandafter\ifx\csname natexlab\endcsname\relax\def\natexlab#1{#1}\fi
\providecommand{\url}[1]{\href{#1}{#1}}
\providecommand{\dodoi}[1]{doi:~\href{http://doi.org/#1}{\nolinkurl{#1}}}
\providecommand{\doeprint}[1]{\href{http://ascl.net/#1}{\nolinkurl{http://ascl.net/#1}}}
\providecommand{\doarXiv}[1]{\href{https://arxiv.org/abs/#1}{\nolinkurl{https://arxiv.org/abs/#1}}}

\bibitem[{{Albano} {et~al.}(2010){Albano}, {Turolla}, {Israel}, {Zane}, {Nobili}, \& {Stella}}]{Albano2010}
{Albano}, A., {Turolla}, R., {Israel}, G.~L., {et~al.} 2010, \apj, 722, 788, \dodoi{10.1088/0004-637X/722/1/788}

\bibitem[{{Alcock} {et~al.}(1986){Alcock}, {Farhi}, \& {Olinto}}]{Alcock1986}
{Alcock}, C., {Farhi}, E., \& {Olinto}, A. 1986, \apj, 310, 261, \dodoi{10.1086/164679}

\bibitem[{{Archibald} {et~al.}(2008){Archibald}, {Kaspi}, {Livingstone}, \& {McLaughlin}}]{2008ApJ...688..550A}
{Archibald}, A.~M., {Kaspi}, V.~M., {Livingstone}, M.~A., \& {McLaughlin}, M.~A. 2008, \apj, 688, 550, \dodoi{10.1086/591661}

\bibitem[{{Archibald} {et~al.}(2016){Archibald}, {Kaspi}, {Tendulkar}, \& {Scholz}}]{2016ApJ...829L..21A}
{Archibald}, R.~F., {Kaspi}, V.~M., {Tendulkar}, S.~P., \& {Scholz}, P. 2016, \apjl, 829, L21, \dodoi{10.3847/2041-8205/829/1/L21}

\bibitem[{{Baring} \& {Harding}(1998)}]{1998ApJ...507L..55B}
{Baring}, M.~G., \& {Harding}, A.~K. 1998, \apjl, 507, L55, \dodoi{10.1086/311679}

\bibitem[{{Baym} {et~al.}(1971){Baym}, {Pethick}, \& {Sutherland}}]{Baym1971}
{Baym}, G., {Pethick}, C., \& {Sutherland}, P. 1971, \apj, 170, 299, \dodoi{10.1086/151216}

\bibitem[{{Beloborodov}(2009)}]{Beloborodov2009}
{Beloborodov}, A.~M. 2009, \apj, 703, 1044, \dodoi{10.1088/0004-637X/703/1/1044}

\bibitem[{{Beloborodov}(2013)}]{Beloborodov2013}
---. 2013, \apj, 762, 13, \dodoi{10.1088/0004-637X/762/1/13}

\bibitem[{{Bernardini} {et~al.}(2011){Bernardini}, {Perna}, {Gotthelf}, {Israel}, {Rea}, \& {Stella}}]{2011MNRAS.418..638B}
{Bernardini}, F., {Perna}, R., {Gotthelf}, E.~V., {et~al.} 2011, \mnras, 418, 638, \dodoi{10.1111/j.1365-2966.2011.19513.x}

\bibitem[{{Borghese} {et~al.}(2020){Borghese}, {Coti Zelati}, {Rea}, {Esposito}, {Israel}, {Mereghetti}, \& {Tiengo}}]{2020ApJ...902L...2B}
{Borghese}, A., {Coti Zelati}, F., {Rea}, N., {et~al.} 2020, \apjl, 902, L2, \dodoi{10.3847/2041-8213/aba82a}

\bibitem[{{Borghese} {et~al.}(2021){Borghese}, {Rea}, {Turolla}, {Rigoselli}, {Alford}, {Gotthelf}, {Burgay}, {Possenti}, {Zane}, {Coti Zelati}, {Perna}, {Esposito}, {Mereghetti}, {Vigan{\`o}}, {Tiengo}, {G{\"o}tz}, {Ibrahim}, {Israel}, {Pons}, \& {Sathyaprakash}}]{Borghese2021}
{Borghese}, A., {Rea}, N., {Turolla}, R., {et~al.} 2021, \mnras, 504, 5244, \dodoi{10.1093/mnras/stab1236}

\bibitem[{{Caleb} {et~al.}(2022){Caleb}, {Rajwade}, {Desvignes}, {Stappers}, {Lyne}, {Weltevrede}, {Kramer}, {Levin}, \& {Surnis}}]{2022MNRAS.510.1996C}
{Caleb}, M., {Rajwade}, K., {Desvignes}, G., {et~al.} 2022, \mnras, 510, 1996, \dodoi{10.1093/mnras/stab3223}

\bibitem[{{Camilo} {et~al.}(2000){Camilo}, {Kaspi}, {Lyne}, {Manchester}, {Bell}, {D'Amico}, {McKay}, \& {Crawford}}]{2000ApJ...541..367C}
{Camilo}, F., {Kaspi}, V.~M., {Lyne}, A.~G., {et~al.} 2000, \apj, 541, 367, \dodoi{10.1086/309435}

\bibitem[{{Camilo} {et~al.}(2006){Camilo}, {Ransom}, {Halpern}, {Reynolds}, {Helfand}, {Zimmerman}, \& {Sarkissian}}]{2006Natur.442..892C}
{Camilo}, F., {Ransom}, S.~M., {Halpern}, J.~P., {et~al.} 2006, \nat, 442, 892, \dodoi{10.1038/nature04986}

\bibitem[{{Camilo} {et~al.}(2007{\natexlab{a}}){Camilo}, {Reynolds}, {Johnston}, {Halpern}, {Ransom}, \& {van Straten}}]{2007ApJ...659L..37C}
{Camilo}, F., {Reynolds}, J., {Johnston}, S., {et~al.} 2007{\natexlab{a}}, \apjl, 659, L37, \dodoi{10.1086/516630}

\bibitem[{{Camilo} {et~al.}(2007{\natexlab{b}}){Camilo}, {Cognard}, {Ransom}, {Halpern}, {Reynolds}, {Zimmerman}, {Gotthelf}, {Helfand}, {Demorest}, {Theureau}, \& {Backer}}]{2007ApJ...663..497C}
{Camilo}, F., {Cognard}, I., {Ransom}, S.~M., {et~al.} 2007{\natexlab{b}}, \apj, 663, 497, \dodoi{10.1086/518226}

\bibitem[{{Camilo} {et~al.}(2016){Camilo}, {Ransom}, {Halpern}, {Alford}, {Cognard}, {Reynolds}, {Johnston}, {Sarkissian}, \& {van Straten}}]{Camilo2016}
{Camilo}, F., {Ransom}, S.~M., {Halpern}, J.~P., {et~al.} 2016, \apj, 820, 110, \dodoi{10.3847/0004-637X/820/2/110}

\bibitem[{{Chatterjee} {et~al.}(2000){Chatterjee}, {Hernquist}, \& {Narayan}}]{Chatterjee2000}
{Chatterjee}, P., {Hernquist}, L., \& {Narayan}, R. 2000, \apj, 534, 373, \dodoi{10.1086/308748}

\bibitem[{{Cheng} \& {Ruderman}(1977)}]{1977ApJ...216..865C}
{Cheng}, A.~F., \& {Ruderman}, M.~A. 1977, \apj, 216, 865, \dodoi{10.1086/155531}

\bibitem[{{Cheng} \& {Dai}(1998)}]{Cheng1998a}
{Cheng}, K.~S., \& {Dai}, Z.~G. 1998, \prl, 80, 18, \dodoi{10.1103/PhysRevLett.80.18}

\bibitem[{{Cheng} {et~al.}(1998){Cheng}, {Dai}, {Wei}, \& {Lu}}]{Cheng1998b}
{Cheng}, K.~S., {Dai}, Z.~G., {Wei}, D.~M., \& {Lu}, T. 1998, Science, 280, 407, \dodoi{10.1126/science.280.5362.407}

\bibitem[{{Chu} {et~al.}(2021){Chu}, {Ng}, {Kong}, \& {Chang}}]{2021MNRAS.503.1214C}
{Chu}, C.-Y., {Ng}, C.~Y., {Kong}, A. K.~H., \& {Chang}, H.-K. 2021, \mnras, 503, 1214, \dodoi{10.1093/mnras/stab349}

\bibitem[{{Coti Zelati} {et~al.}(2018){Coti Zelati}, {Rea}, {Pons}, {Campana}, \& {Esposito}}]{CotiZelati2018}
{Coti Zelati}, F., {Rea}, N., {Pons}, J.~A., {Campana}, S., \& {Esposito}, P. 2018, \mnras, 474, 961, \dodoi{10.1093/mnras/stx2679}

\bibitem[{{Dai}(2020)}]{Dai2020}
{Dai}, Z.~G. 2020, \apjl, 897, L40, \dodoi{10.3847/2041-8213/aba11b}

\bibitem[{{Dai} \& {Lu}(1998)}]{Dai1998}
{Dai}, Z.~G., \& {Lu}, T. 1998, \prl, 81, 4301, \dodoi{10.1103/PhysRevLett.81.4301}

\bibitem[{{De} {et~al.}(2020){De}, {Ashley}, {Andreoni}, {Kasliwal}, {Soria}, {Srinivasaragavan}, {Cai}, {Delacroix}, {Greffe}, {Hale}, {Hankins}, {Li}, {McKenna}, {Moore}, {Ofek}, {Smith}, {Soon}, {Travouillon}, \& {Zhang}}]{De2020}
{De}, K., {Ashley}, M. C.~B., {Andreoni}, I., {et~al.} 2020, \apjl, 901, L7, \dodoi{10.3847/2041-8213/abb3c5}

\bibitem[{{Dib} \& {Kaspi}(2014)}]{2014ApJ...784...37D}
{Dib}, R., \& {Kaspi}, V.~M. 2014, \apj, 784, 37, \dodoi{10.1088/0004-637X/784/1/37}

\bibitem[{{Dib} {et~al.}(2012){Dib}, {Kaspi}, {Scholz}, \& {Gavriil}}]{2012ApJ...748....3D}
{Dib}, R., {Kaspi}, V.~M., {Scholz}, P., \& {Gavriil}, F.~P. 2012, \apj, 748, 3, \dodoi{10.1088/0004-637X/748/1/3}

\bibitem[{{Duncan} \& {Thompson}(1992)}]{Duncan1992}
{Duncan}, R.~C., \& {Thompson}, C. 1992, \apjl, 392, L9, \dodoi{10.1086/186413}

\bibitem[{{Eie} {et~al.}(2021){Eie}, {Terasawa}, {Akahori}, {Oyama}, {Hirota}, {Yonekura}, {Enoto}, {Sekido}, {Takefuji}, {Misawa}, {Tsuchiya}, {Kisaka}, {Aoki}, \& {Honma}}]{2021PASJ...73.1563E}
{Eie}, S., {Terasawa}, T., {Akahori}, T., {et~al.} 2021, \pasj, 73, 1563, \dodoi{10.1093/pasj/psab098}

\bibitem[{{Esposito} {et~al.}(2021){Esposito}, {Rea}, \& {Israel}}]{Esposito2021}
{Esposito}, P., {Rea}, N., \& {Israel}, G.~L. 2021, in Astrophysics and Space Science Library, Vol. 461, Timing Neutron Stars: Pulsations, Oscillations and Explosions, ed. T.~M. {Belloni}, M.~{M{\'e}ndez}, \& C.~{Zhang}, 97--142, \dodoi{10.1007/978-3-662-62110-3_3}

\bibitem[{{Esposito} {et~al.}(2020){Esposito}, {Rea}, {Borghese}, {Coti Zelati}, {Vigan{\`o}}, {Israel}, {Tiengo}, {Ridolfi}, {Possenti}, {Burgay}, {G{\"o}tz}, {Pintore}, {Stella}, {Dehman}, {Ronchi}, {Campana}, {Garcia-Garcia}, {Graber}, {Mereghetti}, {Perna}, {Rodr{\'\i}guez Castillo}, {Turolla}, \& {Zane}}]{Esposito2020}
{Esposito}, P., {Rea}, N., {Borghese}, A., {et~al.} 2020, \apjl, 896, L30, \dodoi{10.3847/2041-8213/ab9742}

\bibitem[{{Fisher} {et~al.}(2024){Fisher}, {Butterworth}, {Rajwade}, {Stappers}, {Desvignes}, {Karuppusamy}, {Kramer}, {Liu}, {Lyne}, {Mickaliger}, {Shaw}, \& {Weltevrede}}]{2024MNRAS.528.3833F}
{Fisher}, R., {Butterworth}, E.~M., {Rajwade}, K.~M., {et~al.} 2024, \mnras, 528, 3833, \dodoi{10.1093/mnras/stae271}

\bibitem[{{Geng} {et~al.}(2021){Geng}, {Li}, \& {Huang}}]{Geng2021}
{Geng}, J., {Li}, B., \& {Huang}, Y. 2021, The Innovation, 2, 100152, \dodoi{10.1016/j.xinn.2021.100152}

\bibitem[{{Glendenning} \& {Weber}(1992)}]{Glendenning1992}
{Glendenning}, N.~K., \& {Weber}, F. 1992, \apj, 400, 647, \dodoi{10.1086/172026}

\bibitem[{{Goldreich} \& {Julian}(1969)}]{Goldreich1969}
{Goldreich}, P., \& {Julian}, W.~H. 1969, \apj, 157, 869, \dodoi{10.1086/150119}

\bibitem[{{Gotthelf} \& {Halpern}(2007)}]{2007ApJ...664L..35G}
{Gotthelf}, E.~V., \& {Halpern}, J.~P. 2007, \apjl, 664, L35, \dodoi{10.1086/520637}

\bibitem[{{Gotthelf} \& {Halpern}(2009)}]{2009ApJ...695L..35G}
---. 2009, \apjl, 695, L35, \dodoi{10.1088/0004-637X/695/1/L35}

\bibitem[{{Gotthelf} \& {Halpern}(2020)}]{Gotthelf2020}
---. 2020, \apj, 900, 159, \dodoi{10.3847/1538-4357/aba7bc}

\bibitem[{{Gotthelf} {et~al.}(2000){Gotthelf}, {Vasisht}, {Boylan-Kolchin}, \& {Torii}}]{2000ApJ...542L..37G}
{Gotthelf}, E.~V., {Vasisht}, G., {Boylan-Kolchin}, M., \& {Torii}, K. 2000, \apjl, 542, L37, \dodoi{10.1086/312923}

\bibitem[{{Gotthelf} {et~al.}(2019){Gotthelf}, {Halpern}, {Alford}, {Mihara}, {Negoro}, {Kawai}, {Dai}, {Lower}, {Johnston}, {Bailes}, {Os{\l}owski}, {Camilo}, {Miyasaka}, \& {Madsen}}]{2019ApJ...874L..25G}
{Gotthelf}, E.~V., {Halpern}, J.~P., {Alford}, J.~A.~J., {et~al.} 2019, \apjl, 874, L25, \dodoi{10.3847/2041-8213/ab101a}

\bibitem[{{Haensel} {et~al.}(1986){Haensel}, {Zdunik}, \& {Schaefer}}]{Haensel1986}
{Haensel}, P., {Zdunik}, J.~L., \& {Schaefer}, R. 1986, \aap, 160, 121

\bibitem[{{Halpern} \& {Gotthelf}(2010)}]{Halpern2010}
{Halpern}, J.~P., \& {Gotthelf}, E.~V. 2010, \apj, 709, 436, \dodoi{10.1088/0004-637X/709/1/436}

\bibitem[{{Halpern} {et~al.}(2005){Halpern}, {Gotthelf}, {Becker}, {Helfand}, \& {White}}]{2005ApJ...632L..29H}
{Halpern}, J.~P., {Gotthelf}, E.~V., {Becker}, R.~H., {Helfand}, D.~J., \& {White}, R.~L. 2005, \apjl, 632, L29, \dodoi{10.1086/497537}

\bibitem[{{Halpern} {et~al.}(2012){Halpern}, {Gotthelf}, \& {Camilo}}]{Halpern2012}
{Halpern}, J.~P., {Gotthelf}, E.~V., \& {Camilo}, F. 2012, \apjl, 753, L14, \dodoi{10.1088/2041-8205/753/1/L14}

\bibitem[{{Harding} {et~al.}(1999){Harding}, {Contopoulos}, \& {Kazanas}}]{1999ApJ...525L.125H}
{Harding}, A.~K., {Contopoulos}, I., \& {Kazanas}, D. 1999, \apjl, 525, L125, \dodoi{10.1086/312339}

\bibitem[{{Haskell} {et~al.}(2006){Haskell}, {Jones}, \& {Andersson}}]{Haskell2006}
{Haskell}, B., {Jones}, D.~I., \& {Andersson}, N. 2006, \mnras, 373, 1423, \dodoi{10.1111/j.1365-2966.2006.10998.x}

\bibitem[{{Hobbs} {et~al.}(2004){Hobbs}, {Lyne}, {Kramer}, {Martin}, \& {Jordan}}]{2004MNRAS.353.1311H}
{Hobbs}, G., {Lyne}, A.~G., {Kramer}, M., {Martin}, C.~E., \& {Jordan}, C. 2004, \mnras, 353, 1311, \dodoi{10.1111/j.1365-2966.2004.08157.x}

\bibitem[{{Hu} {et~al.}(2023){Hu}, {Kuiper}, {Harding}, {Younes}, {Blumer}, {Ho}, {Enoto}, {Espinoza}, \& {Gendreau}}]{2023ApJ...952..120H}
{Hu}, C.-P., {Kuiper}, L., {Harding}, A.~K., {et~al.} 2023, \apj, 952, 120, \dodoi{10.3847/1538-4357/acd850}

\bibitem[{{Huang} \& {Lu}(1997)}]{Huang1997}
{Huang}, Y.~F., \& {Lu}, T. 1997, \aap, 325, 189

\bibitem[{{Huang} {et~al.}(2023){Huang}, {Yan}, {Shen}, {Tong}, {Yuan}, {Lin}, {Zhao}, {Wu}, {Liu}, {Wang}, \& {Wang}}]{2023ApJ...956...93H}
{Huang}, Z.-P., {Yan}, Z., {Shen}, Z.-Q., {et~al.} 2023, \apj, 956, 93, \dodoi{10.3847/1538-4357/acf193}

\bibitem[{{Hughes} {et~al.}(1981){Hughes}, {Harten}, \& {van den Bergh}}]{1981ApJ...246L.127H}
{Hughes}, V.~A., {Harten}, R.~H., \& {van den Bergh}, S. 1981, \apjl, 246, L127, \dodoi{10.1086/183568}

\bibitem[{{Ibrahim} {et~al.}(2004){Ibrahim}, {Markwardt}, {Swank}, {Ransom}, {Roberts}, {Kaspi}, {Woods}, {Safi-Harb}, {Balman}, {Parke}, {Kouveliotou}, {Hurley}, \& {Cline}}]{2004ApJ...609L..21I}
{Ibrahim}, A.~I., {Markwardt}, C.~B., {Swank}, J.~H., {et~al.} 2004, \apjl, 609, L21, \dodoi{10.1086/422636}

\bibitem[{{Ibrahim} {et~al.}(2024){Ibrahim}, {Borghese}, {Coti Zelati}, {Parent}, {Marino}, {Ould-Boukattine}, {Rea}, {Ascenzi}, {Pacholski}, {Mereghetti}, {Israel}, {Tiengo}, {Possenti}, {Burgay}, {Turolla}, {Zane}, {Esposito}, {G{\"o}tz}, {Campana}, {Kirsten}, {Gawro{\'n}ski}, \& {Hessels}}]{2024ApJ...965...87I}
{Ibrahim}, A.~Y., {Borghese}, A., {Coti Zelati}, F., {et~al.} 2024, \apj, 965, 87, \dodoi{10.3847/1538-4357/ad293b}

\bibitem[{{Israel} {et~al.}(2021){Israel}, {Burgay}, {Rea}, {Esposito}, {Possenti}, {Dall'Osso}, {Stella}, {Pilia}, {Tiengo}, {Ridnaia}, {Lien}, {Frederiks}, \& {Bernardini}}]{2021ApJ...907....7I}
{Israel}, G.~L., {Burgay}, M., {Rea}, N., {et~al.} 2021, \apj, 907, 7, \dodoi{10.3847/1538-4357/abca95}

\bibitem[{{Iwasawa} {et~al.}(1992){Iwasawa}, {Koyama}, \& {Halpern}}]{1992PASJ...44....9I}
{Iwasawa}, K., {Koyama}, K., \& {Halpern}, J.~P. 1992, \pasj, 44, 9

\bibitem[{{Jessner} {et~al.}(2001){Jessner}, {Lesch}, \& {Kunzl}}]{2001ApJ...547..959J}
{Jessner}, A., {Lesch}, H., \& {Kunzl}, T. 2001, \apj, 547, 959, \dodoi{10.1086/318379}

\bibitem[{{Kaspi} \& {Beloborodov}(2017)}]{Kaspi2017}
{Kaspi}, V.~M., \& {Beloborodov}, A.~M. 2017, \araa, 55, 261, \dodoi{10.1146/annurev-astro-081915-023329}

\bibitem[{{Kaspi} \& {McLaughlin}(2005)}]{2005ApJ...618L..41K}
{Kaspi}, V.~M., \& {McLaughlin}, M.~A. 2005, \apjl, 618, L41, \dodoi{10.1086/427628}

\bibitem[{{Katz}(2020)}]{Katz2020}
{Katz}, J.~I. 2020, \mnras, 499, 2319, \dodoi{10.1093/mnras/staa3042}

\bibitem[{{Levin} {et~al.}(2019){Levin}, {Lyne}, {Desvignes}, {Eatough}, {Karuppusamy}, {Kramer}, {Mickaliger}, {Stappers}, \& {Weltevrede}}]{2019MNRAS.488.5251L}
{Levin}, L., {Lyne}, A.~G., {Desvignes}, G., {et~al.} 2019, \mnras, 488, 5251, \dodoi{10.1093/mnras/stz2074}

\bibitem[{{Liu} {et~al.}(2023){Liu}, {Gao}, {Li}, {Dohi}, {Wang}, {L{\"u}}, \& {Xu}}]{2023MNRAS.525.2054L}
{Liu}, H., {Gao}, Y., {Li}, Z., {et~al.} 2023, \mnras, 525, 2054, \dodoi{10.1093/mnras/stad2424}

\bibitem[{{Margalit} {et~al.}(2020){Margalit}, {Beniamini}, {Sridhar}, \& {Metzger}}]{Margalit2020}
{Margalit}, B., {Beniamini}, P., {Sridhar}, N., \& {Metzger}, B.~D. 2020, \apjl, 899, L27, \dodoi{10.3847/2041-8213/abac57}

\bibitem[{{Marsden} {et~al.}(2001){Marsden}, {Lingenfelter}, {Rothschild}, \& {Higdon}}]{2001ApJ...550..397M}
{Marsden}, D., {Lingenfelter}, R.~E., {Rothschild}, R.~E., \& {Higdon}, J.~C. 2001, \apj, 550, 397, \dodoi{10.1086/319701}

\bibitem[{{McLaughlin} {et~al.}(2003){McLaughlin}, {Stairs}, {Kaspi}, {Lorimer}, {Kramer}, {Lyne}, {Manchester}, {Camilo}, {Hobbs}, {Possenti}, {D'Amico}, \& {Faulkner}}]{2003ApJ...591L.135M}
{McLaughlin}, M.~A., {Stairs}, I.~H., {Kaspi}, V.~M., {et~al.} 2003, \apjl, 591, L135, \dodoi{10.1086/377212}

\bibitem[{{Mereghetti} {et~al.}(2006){Mereghetti}, {Esposito}, {Tiengo}, {Zane}, {Turolla}, {Stella}, {Israel}, {G{\"o}tz}, \& {Feroci}}]{2006ApJ...653.1423M}
{Mereghetti}, S., {Esposito}, P., {Tiengo}, A., {et~al.} 2006, \apj, 653, 1423, \dodoi{10.1086/508682}

\bibitem[{{Michel}(1975)}]{1975ApJ...198..683M}
{Michel}, F.~C. 1975, \apj, 198, 683, \dodoi{10.1086/153646}

\bibitem[{{Olausen} \& {Kaspi}(2014)}]{2014ApJS...212..6O}
{Olausen}, S.~A., \& {Kaspi}, V.~M. 2014, \apjs, 212, 6, \dodoi{10.1088/0067-0049/212/1/6}

\bibitem[{{Perna} \& {Gotthelf}(2008)}]{Perna2008}
{Perna}, R., \& {Gotthelf}, E.~V. 2008, \apj, 681, 522, \dodoi{10.1086/588211}

\bibitem[{{Pines} \& {Alpar}(1985)}]{Pines1985}
{Pines}, D., \& {Alpar}, M.~A. 1985, \nat, 316, 27, \dodoi{10.1038/316027a0}

\bibitem[{{Pintore} {et~al.}(2019){Pintore}, {Mereghetti}, {Esposito}, {Turolla}, {Tiengo}, {Rea}, {Bernardini}, \& {Israel}}]{Pintore2019}
{Pintore}, F., {Mereghetti}, S., {Esposito}, P., {et~al.} 2019, \mnras, 483, 3832, \dodoi{10.1093/mnras/sty3378}

\bibitem[{{Qiao} {et~al.}(2010){Qiao}, {Xu}, \& {Du}}]{2010arXiv1005.3911Q}
{Qiao}, G.~J., {Xu}, R.~X., \& {Du}, Y.~J. 2010, arXiv e-prints, arXiv:1005.3911, \dodoi{10.48550/arXiv.1005.3911}

\bibitem[{{Rea} {et~al.}(2016){Rea}, {Borghese}, {Esposito}, {Coti Zelati}, {Bachetti}, {Israel}, \& {De Luca}}]{2016ApJ...828L..13R}
{Rea}, N., {Borghese}, A., {Esposito}, P., {et~al.} 2016, \apjl, 828, L13, \dodoi{10.3847/2041-8205/828/1/L13}

\bibitem[{{Rea} {et~al.}(2013){Rea}, {Israel}, {Pons}, {Turolla}, {Vigan{\`o}}, {Zane}, {Esposito}, {Perna}, {Papitto}, {Terreran}, {Tiengo}, {Salvetti}, {Girart}, {Palau}, {Possenti}, {Burgay}, {G{\"o}{\u{g}}{\"u}{\c{s}}}, {Caliandro}, {Kouveliotou}, {G{\"o}tz}, {Mignani}, {Ratti}, \& {Stella}}]{2013ApJ...770...65R}
{Rea}, N., {Israel}, G.~L., {Pons}, J.~A., {et~al.} 2013, \apj, 770, 65, \dodoi{10.1088/0004-637X/770/1/65}

\bibitem[{{Ruderman} \& {Sutherland}(1975)}]{Ruderman1975}
{Ruderman}, M.~A., \& {Sutherland}, P.~G. 1975, \apj, 196, 51, \dodoi{10.1086/153393}

\bibitem[{{Shi} \& {Xu}(2003)}]{Shi2003}
{Shi}, Y., \& {Xu}, R.~X. 2003, \apjl, 596, L75, \dodoi{10.1086/379111}

\bibitem[{{Suvorov}(2023)}]{2023MNRAS.523.4089S}
{Suvorov}, A.~G. 2023, \mnras, 523, 4089, \dodoi{10.1093/mnras/stad1672}

\bibitem[{{Thompson} \& {Duncan}(1995)}]{Thompson1995}
{Thompson}, C., \& {Duncan}, R.~C. 1995, \mnras, 275, 255, \dodoi{10.1093/mnras/275.2.255}

\bibitem[{{Tong} {et~al.}(2013){Tong}, {Xu}, {Song}, \& {Qiao}}]{2013ApJ...768..144T}
{Tong}, H., {Xu}, R.~X., {Song}, L.~M., \& {Qiao}, G.~J. 2013, \apj, 768, 144, \dodoi{10.1088/0004-637X/768/2/144}

\bibitem[{{Torne} {et~al.}(2020){Torne}, {Mac{\'\i}as-P{\'e}rez}, {Ladjelate}, {Ritacco}, {S{\'a}nchez-Portal}, {Berta}, {Paubert}, {Calvo}, {Desvignes}, {Karuppusamy}, {Navarro}, {John}, {S{\'a}nchez}, {Pe{\~n}alver}, {Kramer}, \& {Schuster}}]{Torne2020}
{Torne}, P., {Mac{\'\i}as-P{\'e}rez}, J., {Ladjelate}, B., {et~al.} 2020, \aap, 640, L2, \dodoi{10.1051/0004-6361/202038504}

\bibitem[{{Usov}(2001)}]{Usov2001}
{Usov}, V.~V. 2001, \prl, 87, 021101, \dodoi{10.1103/PhysRevLett.87.021101}

\bibitem[{{Vrba} {et~al.}(2000){Vrba}, {Henden}, {Luginbuhl}, {Guetter}, {Hartmann}, \& {Klose}}]{2000ApJ...533L..17V}
{Vrba}, F.~J., {Henden}, A.~A., {Luginbuhl}, C.~B., {et~al.} 2000, \apjl, 533, L17, \dodoi{10.1086/312602}

\bibitem[{{Wang}(2020)}]{Wang2020}
{Wang}, J.-S. 2020, \apj, 900, 172, \dodoi{10.3847/1538-4357/aba955}

\bibitem[{{Wang} {et~al.}(2018){Wang}, {Luo}, {Yue}, {Chen}, {Lee}, \& {Xu}}]{2018ApJ...852..140W}
{Wang}, W., {Luo}, R., {Yue}, H., {et~al.} 2018, \apj, 852, 140, \dodoi{10.3847/1538-4357/aaa025}

\bibitem[{{Woods} {et~al.}(2007){Woods}, {Kouveliotou}, {Finger}, {G{\"o}{\v{g}}{\"u}{\c{s}}}, {Wilson}, {Patel}, {Hurley}, \& {Swank}}]{2007ApJ...654..470W}
{Woods}, P.~M., {Kouveliotou}, C., {Finger}, M.~H., {et~al.} 2007, \apj, 654, 470, \dodoi{10.1086/507459}

\bibitem[{{Xu}(2002)}]{2002ApJ...570L..65X}
{Xu}, R.~X. 2002, \apjl, 570, L65, \dodoi{10.1086/340993}

\bibitem[{{Xu}(2003)}]{2003ApJ...596L..59X}
---. 2003, \apjl, 596, L59, \dodoi{10.1086/379209}

\bibitem[{{Xu} \& {Busse}(2001)}]{Xu2001}
{Xu}, R.~X., \& {Busse}, F.~H. 2001, \aap, 371, 963, \dodoi{10.1051/0004-6361:20010450}

\bibitem[{{Xu} {et~al.}(1999){Xu}, {Qiao}, \& {Zhang}}]{1999ApJ...522L.109X}
{Xu}, R.~X., {Qiao}, G.~J., \& {Zhang}, B. 1999, \apjl, 522, L109, \dodoi{10.1086/312226}

\bibitem[{{Xu} {et~al.}(2006){Xu}, {Tao}, \& {Yang}}]{Xu2006}
{Xu}, R.~X., {Tao}, D.~J., \& {Yang}, Y. 2006, \mnras, 373, L85, \dodoi{10.1111/j.1745-3933.2006.00248.x}

\bibitem[{{Zhang} \& {Harding}(2000)}]{2000ApJ...535L..51Z}
{Zhang}, B., \& {Harding}, A.~K. 2000, \apjl, 535, L51, \dodoi{10.1086/312694}

\bibitem[{{Zhu} {et~al.}(2023){Zhu}, {Xu}, {Zhou}, {Lin}, {Wang}, {Wang}, {Zhang}, {Niu}, {Chen}, {Li}, {Meng}, {Lee}, {Zhang}, {Feng}, {Ge}, {G{\"o}{\u{g}}{\"u}{\c{s}}}, {Guan}, {Han}, {Jiang}, {Jiang}, {Kouveliotou}, {Li}, {Miao}, {Miao}, {Men}, {Niu}, {Wang}, {Wang}, {Xu}, {Xu}, {Xue}, {Yang}, {Yu}, {Yuan}, {Yue}, {Zhang}, \& {Zhang}}]{2023SciA....9F6198Z}
{Zhu}, W., {Xu}, H., {Zhou}, D., {et~al.} 2023, Science Advances, 9, eadf6198, \dodoi{10.1126/sciadv.adf6198}

\end{thebibliography}

\bibliographystyle{aasjournal}

\end{document}